\documentclass[11pt]{article}

\usepackage[usenames,dvipsnames]{xcolor}
\usepackage{ctable,colonequals}
\usepackage{amsfonts}
\usepackage{amsmath}
\usepackage{amssymb}
\usepackage{subfigure}
\usepackage{mathtools}
\usepackage{multirow,natbib}

\newcommand{\real}{I\hspace{-0.9mm}R}

\def\bzero{\boldsymbol{0}}

\def\bx{\boldsymbol{x}}

\def\bv{\boldsymbol{v}}
\def\bz{\boldsymbol{z}}

\def\bX{\boldsymbol{X}}
\def\bY{\boldsymbol{Y}}

\def\balpha{\boldsymbol{\alpha}}

\def\bmu{\boldsymbol{\mu}}
\def\bpi{\boldsymbol{\pi}}
\def\blambda{\boldsymbol{\lambda}}
\def\brho{\boldsymbol{\rho}}
\def\btheta{\boldsymbol{\theta}}
\def\bvartheta{\boldsymbol{\vartheta}}

\def\bSigma{\boldsymbol{\Sigma}}

\begin{document}
\title{Asymmetric Clusters and Outliers: Mixtures of Multivariate Contaminated Shifted Asymmetric Laplace Distributions}
\author{Katherine Morris$^*$, Antonio Punzo$^{**}$, Paul D. McNicholas$^{\dagger}$ and Ryan P. Browne$^{\dagger\dagger}$}
\date{\small $^*$Dept.\ of Mathematics \& Statistics, University of Guelph, Guelph, Ontario, Canada.\\
$^{**}$Dept.\ of Economics and Business, University of Catania, Italy.\\
$^{\dagger}$Dept.\ of Mathematics \& Statistics, McMaster University, Hamilton, Ontario, Canada.\\
$^{\dagger\dagger}$Dept.\ of Statistics and Actuarial Science, University of Waterloo, Ontario, Canada.}

\maketitle

\begin{abstract}
Mixtures of multivariate contaminated shifted asymmetric Laplace distributions are developed for handling asymmetric clusters in the presence of outliers (also referred to as bad points herein).
In addition to the parameters of the related non-contaminated mixture, for each (asymmetric) cluster, our model has one parameter controlling the proportion of outliers and one specifying the degree of contamination. 
Crucially, these parameters do not have to be specified \textit{a~priori}, adding a flexibility to our approach that is absent from other approaches such as trimming.
Moreover, each observation is given a posterior probability of belonging to a particular cluster, and of being an outlier or not; advantageously, this allows for the automatic detection of outliers.
An expectation-conditional maximization algorithm is outlined for parameter estimation and various implementation issues are discussed.
The behaviour of the proposed model is investigated, and compared with well-established finite mixtures, on artificial and real data.\\[-10pt]

\noindent\textbf{Keywords}: Outlier detection, mixture models; model-based clustering; shifted asymmetric Laplace distribution; contaminated normal distribution.
\end{abstract}

\section{Introduction}
\label{sec:introduction}

Clustering algorithms based on probability models are a popular choice for exploring complex data structures.
The model-based approach assumes that data are generated by a finite mixture of probability distributions. 
A $p$-dimensional random vector $\bX$ is said to arise from a parametric finite mixture of $G$ distributions if its probability density function (pdf) is a convex linear combination of pdfs, i.e.,  
\begin{equation*}
p\left(\bx |\bvartheta\right)=\sum_{g=1}^G\pi_g f\left(\bx| \btheta_g\right),
%\label{eq:finitemixture}
\end{equation*}
where $\bvartheta=(\pi_1, \ldots, \pi_G, \btheta_1,\ldots, \btheta_G)$ is the overall parameter vector, $\pi_g\in\left(0,1\right]$ is the mixing proportion, so that $\sum_{g=1}^G\pi_g=1$, and $f\left(\bx| \btheta_g\right)$ is the component pdf, $g=1,\ldots,G$.

In the mixture family, finite mixtures of multivariate normal distributions have received considerable attention because of their computational and theoretical convenience by assuming, in most cases, that each mixture component represents a cluster (or group) within the original data (see, e.g., \citealp{Fral:Raft:Mode:2002} and \citealp{mcnicholas16b}). 
If a component pdf in the mixture is associated with a cluster, as we assume in this paper \citep[see][Section~9.1]{mcnicholas16a}, then normal mixture components imply elliptically symmetric (or elliptically contoured) clusters, which is rather restrictive.
Other examples of mixtures implying elliptically symmetric clusters are those with multivariate $t$ \citep{Peel:McLa:Robu:2000}, multivariate power exponential (\citealp{Zhan:Lian:Robu:2010} and \citealp{Dang:Brow:McNi:Mixt:2015}), and multivariate leptokurtic-normal \citep{Bagn:Punz:Zoia:Them:2016} components.  
One way to overcome such a restrictiveness is to argue that asymmetric clusters can be approximated quite well by a mixture of several elliptically symmetric densities like the normal one (\citealp{Dasg:Raft:Dete:1998}, \citealp{McLa:Peel:Fini:2000}, and \citealp{Titt:Smit:Mako:stat:1985}). 
While this can be very helpful for modelling purposes, it might be misleading when dealing with clustering applications because one group may be represented by more than one component just because it has, in fact, an asymmetric density \citep[see][for an interesting example]{franczak14}. 
One possible approach to dealing with asymmetric groups consists of considering transformations so as to make the components as elliptical and symmetric as possible, and then fitting mixtures of elliptically symmetric (usually normal) distributions (\citealp{Scho:Scho:Skew:1988} and \citealp{Guti:Carr:Wang:Lee:Tayl:Anal:1995}). 
Although such a treatment is very convenient, the achievement of joint elliptical symmetry is rarely satisfied and the transformed variables become more difficult to interpret. 
Instead of applying transformations, there is a growing interest in proposing mixture models where the component distributions are skewed. 
Examples in this direction are mixtures of multivariate skew-normal (SN) distributions (\citealp{Lin:Maxi:2009} and \citealp{pyne09}), multivariate shifted asymmetric Laplace (SAL) distributions \citep{franczak14}, and other approaches \citep[e.g.,][]{murray17,tang18}.
The mixture of SAL distributions approach has the advantage to simplify the computational effort required by the EM algorithm to fit the model. 

However real data, in addition to being characterized by underlying asymmetric clusters, are often ``contaminated'' by outliers or otherwise ``bad'' points. 
The use of ``bad'' in this sense is by analogy with \citet{Aitk:Wils:Mixt:1980}, and refers to points that have a deleterious effect on parameter estimation, including the mixing proportions, as discussed, for example, in \citet{Gall:Ritt:Trim:2009}. 
Thus an important practical application is the development of methods capable of detecting bad points and performing robust parameter estimation when they are present. 
Examples of mixture models coping with such issues are: mixtures of multivariate skew-$t$ distributions (see, e.g., \citealp{wang09}, \citealp{lin10}, \citealp{Lee:McLa:Fine:2014}, \citealp{vrbik12,vrbik14}, and \citealp{murray14b,murray17b}), mixtures of multivariate $t$-distributions with the Box-Cox transformation \citep{Lo:Gott:Flex:2012}, mixtures of multivariate normal inverse Gaussian distributions \citep{karlis09,subedi14,ohagan16}, mixtures of multivariate skew-slash distributions \citep{cabral12}, mixtures of multivariate generalized hyperbolic distributions \citep{browne15}, scale mixtures of multivariate skew-normal distributions (\citealp{Bass:Lach:Cabr:Ghos:Robu:2010} and \citealp{DaSi:Bolf:Lach:Skew:2011}), variance-gamma distributions \citep{smcnicholas17}, hidden truncation hyperbolic distributions \citep{murray17}, and joint generalized hyperbolic distributions \citep{tang18}. 
However, in the methods cited, there is no automatic way of detecting outliers unless one defines some subjective/exogenous trimming rule. 

To overcome this drawback, in Section~\ref{sec:CSAL mixture} we propose to ``contaminate'' the components (clusters) of the multivariate SAL mixture to accommodate outliers and to allow for their automatic detection (see Section~\ref{subsec:Automatic detection of bad points}).
\textcolor{black}{Contamination is introduced by substituting each SAL cluster with a mixture of two SAL pdfs with the same mode and proportional covariance matrices; this is in line with the approach considered by \citet{Punz:McNi:Robu:2016} to define mixtures of multivariate contaminated normal distributions; see also \citet{Maru:Punz:Mode:2017} and \citet{Mazz:Punz:Mixt:2018}. 
According to our contamination scheme, the unimodality of the cluster distribution is preserved \citep{berger1986robust} and its tails are made heavier to accommodate the occurrence of bad points.}
The choice of working with clusters having a unimodal distribution is justified, but above all natural, if one considers that the most striking feature of a mixture of distributions is often that of multimodality (\citealp{Bagn:Punz:Fine:2013} and \citealp{Punz:Bagn:Maru:Comp:2017}), and a single cluster is most naturally characterized by a unimodal pdf \citep[see][for further discussion]{mcnicholas16a}.
Indeed, as highlighted in \citet{Titt:Smit:Mako:stat:1985} and \citet{McBa:mixt:1988}, many papers in applied fields talk not in terms of mixtures but of multimodal distributions; examples are the articles of \citet{Murp:onec:1964} and \citet{BSCS:bimo:1983} referring to bimodality rather than to mixtures of two distributions. 

After a CSAL mixture is fitted on the available data, each observation can be first assigned to one of the clusters, by means of maximum \textit{a~posteriori} probabilities, and then classified as good or bad, and this is a significant advantage, as explained in Section~\ref{subsec:Automatic detection of bad points}.
Moreover, bad points are automatically down-weighted in the estimation of the parameters of the nested SAL mixture.
Thus, we have a model for simultaneous robust clustering, in the presence of asymmetric clusters, and detection of bad points.

\textcolor{black}{Note that the mixture of multivariate skew-contaminated normal (SCN) distributions, introduced by \citet{cabral12}, is able of coping with skewed clusters under the occurrence of outliers that, eventually, can be also automatically detected (even if the authors do not refer to this possibility).
However, although each SCN distribution is defined as a mixture of two SN distributions, the good and bad SN components of this mixture have different modes (see \citealp{Lach:Ghos:Arel:Like:2010} and \citealp{Lach:labr:Mult:2014} for details).
In this regard note that, although the mode of the SN distribution is unique (\citealp{Azza:Thes:2005} and \citealp[][p.~126]{Azza:Capi:TheS:2014}), there is no analytic expression for it \citep[][p.~140]{Azza:Capi:TheS:2014}.    
Hence, the SCN mixture cannot be considered as fitting within our contamination scheme.}

For maximum likelihood parameter estimation of the CSAL mixture proposed herein, an expectation-conditional maximization (ECM) algorithm \citep{Meng:Rubin:Maxi:1993} is developed (Section~\ref{sec:Parameter estimation}).
Further computational and operational aspects are discussed in Section~\ref{sec:Further aspects}.
Section~\ref{sec:Applications to artificial and real data} investigates the performance of our mixture, in comparison with mixtures of some well-established multivariate elliptically countered and skewed distributions, on artificial and real data.
Section~\ref{sec:conclusion} provides the conclusion and suggestions for future work. 
All computational work herein was carried out using \textsf{R} \citep{R:2017}.
 
\section{Mixtures of contaminated shifted asymmetric Laplace distributions}
\label{sec:CSAL mixture}

\textcolor{black}{In this section, we propose to conveniently modify the finite mixture of multivariate shifted asymmetric Laplace (SAL) distributions, proposed by \citet{franczak14}, for the occurrence of bad points. 
In situations where clusters may be asymmetric, SAL mixtures, and mixtures of skewed distributions in general, are more appropriate than mixtures of elliptically contoured distributions because they do not overfit the data by including additional components to capture skewness (see, e.g., \citealp{Lin:Lee:Yen:Fini:2007} and \citealp{Lin:Maxi:2009}).} 

\citet{franczak14} give the pdf of a component in a multivariate SAL mixture as  
\begin{equation}
f_{\text{SAL}}\left(\bx|\bmu, \bSigma, \balpha\right)=\frac{2 \text{exp}\{\left(\bx-\bmu\right)^{\top} \bSigma^{-1}\balpha\}}{(2\pi)^{p/2}| \bSigma|^{1/2}} \left[\frac{\delta\left(\bx, \bmu | \bSigma\right)}{2+\balpha^{\top} \bSigma^{-1}\balpha}\right]^{\nu/2} K_{\nu}(u),
\label{eq:saldensity}
\end{equation}
where $\bmu$ is the \textcolor{black}{mode (%according to the concept of \textbf{star modality} given in 
see \citealp{Dhar:Joag:Unim:1988} and \citealp[][Chapter~6.6]{Kotz:Kozu:Podg:TheL:2012} for details)}, $\bSigma$ is a scale matrix, $\balpha \in \real^p$ denotes the skewness, $\delta\left(\bx, \bmu | \bSigma\right)=\left(\bx -\bmu\right)^{\top}\bSigma^{-1}(\bx -\bmu)$ is the squared Mahalanobis distance between $\bx$ and $\bmu$, with respect to a scale matrix $\bSigma$, $u=\sqrt{\left(2+\balpha^{\top} \bSigma^{-1} \balpha\right)\delta\left(\bx,\bmu | \bSigma\right)}$, and $K_{\nu}$ is the modified Bessel function of the third kind with index $\nu=(2-p)/2$. 
\textcolor{black}{The mean vector and covariance matrix of the SAL distribution are given by $E\left(\bX\right)=\bmu+\balpha$ and $\text{Cov}\left(\bX\right)=\bSigma+\balpha\balpha^{\top}$, respectively.}
Furthermore, \cite{franczak14} use the fact that a random vector $\bX$ whose density is given in \eqref{eq:saldensity} can be generated through the relationship
\begin{equation}
\bX=\bmu + W\balpha +\sqrt{W}\bY,
\label{eq:salrelation}
\end{equation} 
where $W\sim \mathcal{\text{Exp}}(1)$ and $\bY\sim \mathcal{N}_p(\bzero, \bSigma)$, thus $\bX |W=w \sim \mathcal{N}_p\left(\bmu+w\balpha, w\bSigma\right)$. 

\textcolor{black}{To build the framework for model-based clustering with multivariate contaminated SAL mixtures, we follow the approach adopted by \citet{Punz:McNi:Robu:2016} to define multivariate contaminated normal mixtures.
According to this approach, each of the $G$ clusters (components) of the mixture is itself a mixture of two (multivariate normal) components with equal mode and proportional covariance matrices. 
In each cluster, the component with the lower dispersion fits the good data (``good component'') while the other accommodates the outliers (``bad component'').
To apply this idea to the SAL mixture, 
%With reference to the generic cluster of our multivariate contaminated SAL mixture,
%We apply this idea to each of the $G$ clusters of our multivariate contaminated normal 
%To apply this contamination scheme to the multivariate 
%To apply this idea using the multivariate SAL mixture as reference model for the good data, 
we employ a contamination scheme where, in each cluster, $\bSigma$ and $\balpha$ of the good SAL component are respectively inflated as $\rho\bSigma$ and $\sqrt{\rho}\balpha$ in the bad SAL component, with $\rho>1$ denoting the contamination factor. This means that, for bad points, $$\text{Cov}\left(\bX\right)=\rho\bSigma+\sqrt{\rho}\balpha\sqrt{\rho}\balpha^{\top}=\rho(\bSigma+\balpha\balpha^{\top}),$$ i.e., the covariance matrix for the bad points has been inflated by $\rho$ compared to the covariance matrix for the good points, as in the approach of \citet{Punz:McNi:Robu:2016}. This leads to the contaminated SAL (CSAL) distribution whose pdf is
\begin{equation}
f_{\text{CSAL}}\left(\bx| \bmu,\bSigma,\balpha,\lambda,\rho\right)=
\lambda f_{\text{SAL}}\left(\bx | \bmu, \bSigma, \balpha\right)
+
\left(1-\lambda\right)f_{\text{SAL}}\left(\bx | \bmu, \rho\bSigma, \sqrt{\rho}\balpha\right),
\label{eq:CSAL distribution}
\end{equation}
where $\lambda \in \left(0,1\right)$ denotes the proportion of good points.
%\textcolor{orange}{If $\bX$ has the probability density function in \eqref{eq:CSAL distribution}, then we briefly write $\bX\sim \mathcal{CSAL}_p\left(\bmu,\bSigma,\balpha,\lambda,\rho\right)$.} 
The mode and covariance matrix of the bad component $f_{\text{SAL}}\left(\bx|\bmu,\rho\bSigma,\sqrt{\rho}\balpha\right)$ of the CSAL distribution in \eqref{eq:CSAL distribution} are $\bmu$ and $\rho(\bSigma+\balpha\balpha^{\top})$, respectively.
Hence, the bad component has the same mode $\bmu$, and inflated covariance matrix, with respect to the good component; moreover, $\bmu$ is also the mode of the CSAL distribution.
Thus, our way to contaminate fulfills the requirements of the contamination scheme used by \citet{Punz:McNi:Robu:2016}. 
% Hence, the bad component has the same mode, and inflated covariance matrix, with respect to the good component, and this is in line with the contamination approach of \citet{Punz:McNi:Robu:2016} and \citet{Punz:Mazz:Maru:Fitt:2018}.
%Moreover, $\mu$ is also the mode of the CSAL distribution 
%$\bSigma$ is inflated as $\rho\bSigma$ and $\balpha$ is scaled to $\sqrt{\rho}\balpha$, with $\rho>1$ denoting the contamination factor.  
%\clearpage
%%We follow the approach of \citet{Punz:McNi:Robu:2016} for contaminated normal mixtures \textcolor{black}{(see also AGGIUNGERE LE MIE CITAZIONI)} to build the framework for model-based clustering with contaminated SAL mixtures. 
%To this end, we employ the contamination scheme where $\bSigma$ is scaled to $\rho\bSigma$ and $\balpha$ is scaled to $\sqrt{\rho}\balpha$, with $\rho>1$ denoting the contamination factor. 
We can visualize the result by looking at \figurename~\ref{fig:CSAL contours}; here, the contours from a bivariate SAL pdf with parameters
\begin{equation}
\bmu=\begin{bmatrix*}[r]
    0\\  
		0 \\
\end{bmatrix*}, 
\quad 
\bSigma=\begin{bmatrix*}[r]
 1 & 0.5 \\
 0.5 & 1 \\
\end{bmatrix*} 
\quad \text{and} \quad 
\balpha=\begin{bmatrix*}[r]
    1 \\  
		1 \\
\end{bmatrix*},
\label{eq:generating parameters}
\end{equation}
are compared to the contours from a bivariate CSAL pdf with the same parameters $\bmu$, $\bSigma$ and $\balpha$ given in \eqref{eq:generating parameters}, with proportion of good observations $\lambda=0.8$ and degree of contamination $\rho=5$.
\begin{figure}[!ht]
\centering
\resizebox{0.5\textwidth}{!}{
\includegraphics{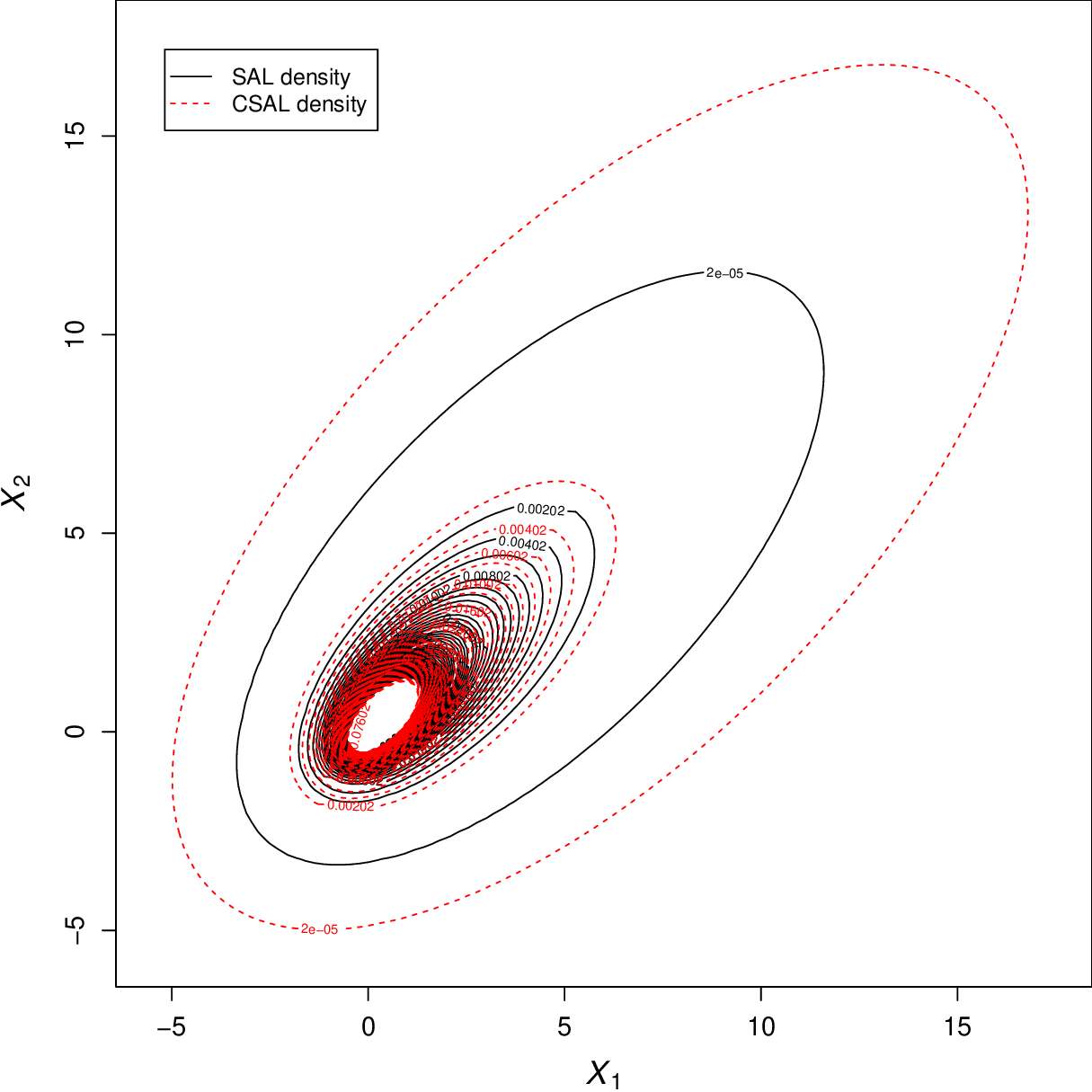} 
}
\caption{Example of contours comparing the SAL (solid black contours) and CSAL (dashed red
contours) distributions.}
\label{fig:CSAL contours}
\end{figure}}
 
\textcolor{black}{The assumption of clusters having a CSAL distribution yields the multivariate CSAL mixture with pdf
\begin{equation}
p_{\text{CSAL}}(\bx| \bvartheta)=\sum_{g=1}^G \pi_g \left[\lambda_g f_{\text{SAL}}\left(\bx | \bmu_g, \bSigma_g, \balpha_g\right)+(1-\lambda_g)f_{\text{SAL}}\left(\bx | \bmu_g, \rho_g\bSigma_g, \sqrt{\rho_g}\balpha_g\right)\right],
\label{eq:CSAL mixture}
\end{equation}
where $\bvartheta=\left(\pi_1,\ldots,\pi_G,\bmu_1,\ldots,\bmu_G,\bSigma_1,\ldots,\bSigma_G,\balpha_1,\ldots,\balpha_G,\lambda_1,\ldots,\lambda_G,\rho_1,\ldots,\rho_G\right)$ contains all the parameters of the model.}

\section{\textcolor{black}{Maximum likelihood estimation: the ECM algorithm}}
\label{sec:Parameter estimation}

Let $\bx_1,\ldots,\bx_n$ be an observed sample from the CSAL mixture model~\eqref{eq:CSAL mixture}.
To find maximum likelihood (ML) estimates for the parameters $\bvartheta$ of this model, we adopt the expectation-conditional maximization (ECM) algorithm \citep{Meng:Rubin:Maxi:1993}.
The ECM algorithm is a variant of the famous expectation-maximization (EM) algorithm \citep{Demp:Lair:Rubi:Maxi:1977}.
The EM algorithm, as well as its variants, are iterative procedures for finding ML estimates when data are incomplete or are treated as being incomplete. 
In our case, as in \citet{Punz:McNi:Robu:2016}, there are two hierarchical sources of incompleteness.
We have a first-step source of incompleteness, the classical one in the use of mixture models, arising from the fact that for each observation $\bx_i$ we do not know its cluster membership; this source is governed by an indicator vector $\bz_i=\left(z_{i1},\ldots,z_{iG}\right)$, where $z_{ig}=1$ if $\bx_i$ comes from cluster $g$ and $z_{ig}=0$ otherwise.
The other source arises from the fact that we do not know whether an observation in group $g$ is good or bad. 
To denote this second source of missing data, we use the indicator vector $\bv_i=\left(v_{i1},\ldots,v_{iG}\right)$ so that $v_{ig}=1$ if ${\bx}_i$ in group~$g$ is good and $v_{ig}=0$ if $\bx_i$ in group $g$ is bad.
The values of $z_{ig}$ and $v_{ig}$ are used for the definition of the following complete-data likelihood 
\begin{equation}
L_c\left(\bvartheta\right)=\prod_{i=1}^n \prod_{g=1}^G \Big\{ \pi_g  [\lambda_g f_{\text{SAL}}(\bx_i | \bmu_g, \bSigma_g,\balpha_g)]^{v_{ig}} [ (1-\lambda_g) f_{\text{SAL}}(\bx_i | \bmu_g, \rho_g\bSigma_g,\sqrt{\rho_g}\balpha_g)]^{(1- v_{ig})} \Big \}^{z_{ig}}.
\label{eq:salcontlike}
\end{equation}
The complete-data log-likelihood corresponding to \eqref{eq:salcontlike} can be written as
\begin{equation}
l_c\left(\bvartheta\right)=l_{c1}\left(\bpi\right)+l_{c2}\left(\blambda\right)+l_{c3}^{\text{good}}\left(\bmu,\bSigma,\balpha\right)+l_{c3}^{\text{bad}}\left(\bmu,\bSigma,\balpha,\brho\right),
\label{eq:complete-data log-likelihood}
\end{equation}
where 
\begin{equation*}
l_{c1}\left(\bpi\right)=\sum_{i=1}^{n}\sum_{g=1}^{G}{z}_{ig}\ln \pi_g,
%\label{eq:complete-data log-likelihood - pi}
\end{equation*}
\begin{equation*}
l_{c2}\left(\blambda\right)=\sum_{i=1}^{n}\sum_{g=1}^{G}z_{ig}\left[v_{ig}\ln \lambda_g+\left(1-v_{ig}\right)\ln \left(1-\lambda_g\right)\right],
%\label{eq:complete-data log-likelihood - lambda}
\end{equation*}
\begin{equation*}
l_{c3}^{\text{good}}\left(\bmu,\bSigma,\balpha\right) = \sum_{i=1}^n \sum_{g=1}^G z_{ig} v_{ig}\log\left[f_{\text{SAL}}(\bx_i | \bmu_g, \bSigma_g,\balpha_g)\right]
%\label{eq:complete-data log-likelihood - good}
\end{equation*}
and 
\begin{equation*}
l_{c3}^{\text{bad}}\left(\bmu,\bSigma,\balpha,\brho\right) = \sum_{i=1}^n \sum_{g=1}^G z_{ig} \left(1-v_{ig}\right)\log\left[f_{\text{SAL}}(\bx_i | \bmu_g, \rho_g\bSigma_g, \sqrt{\rho_g}\balpha_g)\right],
%\label{eq:complete-data log-likelihood - bad}
\end{equation*}
with $\bpi=\left(\pi_1,\ldots,\pi_G\right)$, $\blambda=\left(\lambda_1,\ldots,\lambda_G\right)$, $\bmu=\left(\bmu_1,\ldots,\bmu_G\right)$, $\bSigma=\left(\bSigma_1,\ldots,\bSigma_G\right)$, $\balpha=\left(\balpha_1,\ldots,\balpha_G\right)$ and $\brho=\left(\rho_1,\ldots,\rho_G\right)$.     
Computationally, it is more efficient to use the relationship between the SAL and normal distributions outlined in \eqref{eq:salrelation} to rewrite $l_{c3}^{\text{good}}$ and $l_{c3}^{\text{bad}}$ as
\begin{equation*}
l_{c3}^{\text{good}}\left(\bmu,\bSigma,\balpha\right) = \sum_{i=1}^n \sum_{g=1}^G z_{ig} v_{ig}\log\left[f_{\text{N}}(\bx_i | \bmu_g +w_{ig}\balpha_g, w_{ig}\bSigma_g) f_{\text{Exp}}(w_{ig}|1)\right]
%\label{eq:complete-data log-likelihood - good 2}
\end{equation*}
and
\begin{equation*}
l_{c3}^{\text{bad}}\left(\bmu,\bSigma,\balpha,\brho\right) = \sum_{i=1}^n \sum_{g=1}^G z_{ig} (1-v_{ig})\log\left[f_{\text{N}}(\bx_i | \bmu_g +w_{ig}\sqrt{\rho_g}\balpha_g, w_{ig}\rho_g\bSigma_g) f_{\text{Exp}}(w_{ig}|1)\right],
%\label{eq:complete-data log-likelihood - bad 2}
\end{equation*}
where $f_{\text{Exp}}\left(\cdot|1\right)$ denotes the pdf of an exponential distribution with rate 1, i.e.,~$f_{\text{Exp}}\left(w_{ig}|1\right)=e^{-w_{ig}}$, $w_{ig}>0$.

The ECM algorithm iterates between three steps, one E-step and two CM-steps, until convergence. 
The only difference from the EM algorithm is that each M-step is replaced by two simpler CM-steps. 
They arise from the partition $\bvartheta=\left(\bvartheta_1,\bvartheta_2\right)$, where $\bvartheta_1=\left(\bpi,\bmu,\bSigma,\balpha,\blambda\right)$ and $\bvartheta_2=\brho$.
The three steps of the ECM algorithm, for the generic $\left(r+1\right)$th iteration, $r=1,2,\ldots$, are detailed below.

%%%%%%%%%%
% E-step %
%%%%%%%%%%

\subsection{E-step}
\label{subsec:E-step}

The E-step requires the calculation of $Q\left(\bvartheta\right)=E[l_c\left(\bvartheta\right)|\bx_1,\ldots,\bx_n, \bvartheta^{(r)}]$, the conditional expectation of $l_c\left(\bvartheta\right)$ given the observed data $\bx_1,\ldots,\bx_n$, using the current fit $\bvartheta^{(r)}$ for $\bvartheta$.
According to the decomposition of $l_c\left(\bvartheta\right)$ in \eqref{eq:complete-data log-likelihood}, the $Q$ function can be written as
\begin{equation*}
Q\left(\bvartheta\right)=
Q_1\left(\bpi\right)
+
Q_2\left(\blambda\right)
+
Q_3^{\text{good}}\left(\bmu,\bSigma,\balpha\right)
+
Q_3^{\text{bad}}\left(\bmu,\bSigma,\balpha,\brho\right),
%\label{eq:expected complete-data log-likelihood}
\end{equation*}
where
\begin{equation}
Q_1\left(\bpi\right)=\sum_{g=1}^G n_g^{(r)}\ln \pi_g,
\label{eq:expected complete-data log-likelihood - pi}
\end{equation}
\begin{equation}
Q_2\left(\blambda\right)=\sum_{i=1}^{n}\sum_{g=1}^G z_{ig}^{(r)}\left[v_{ig}^{(r)}\ln \lambda_g+\left(1-v_{ig}^{(r)}\right)\ln \left(1-\lambda_g\right)\right],
\label{eq:expected complete-data log-likelihood - lambda}
\end{equation}
\begin{align}
Q_3^{\text{good}}\left(\bmu,\bSigma,\balpha\right)=&
- \frac{np}{2}\ln\left(2\pi\right) 
- \frac{1}{2}\sum_{g=1}^G n_{g,\text{good}}^{(r)} \ln\left|\bSigma_g\right|
- \frac{p}{2}\sum_{i=1}^n\sum_{g=1}^G z_{ig}^{(r)}v_{ig}^{(r)} E_{3ig}^{(r)}
\nonumber\\
&- \frac{1}{2}\sum_{i=1}^n\sum_{g=1}^G z_{ig}^{(r)}v_{ig}^{(r)} E_{2ig}^{(r)} \left(\bx_i-\bmu_g\right)^{\top}\bSigma_g^{-1}  \left(\bx_i-\bmu_g\right) 
+ \sum_{i=1}^n\sum_{g=1}^G z_{ig}^{(r)}v_{ig}^{(r)} \left(\bx_i-\bmu_g\right)^{\top} \bSigma_g^{-1} \balpha_g
\nonumber\\
&
- \frac{1}{2}\sum_{i=1}^n\sum_{g=1}^G z_{ig}^{(r)}v_{ig}^{(r)} E_{1ig}^{(r)} \balpha_g^{\top}\bSigma_g^{-1}\balpha_g
- \sum_{i=1}^n\sum_{g=1}^G z_{ig}^{(r)}v_{ig}^{(r)} E_{1ig}^{(r)} 
\label{eq:expected complete-data log-likelihood - good}
\end{align}
and
\begin{align}
Q_4^{\text{bad}}\left(\bmu,\bSigma,\balpha,\brho\right)=&
- \frac{np}{2}\ln\left(2\pi\right) 
- \frac{1}{2}\sum_{g=1}^G n_{g,\text{bad}}^{(r)} \ln\left|\bSigma_g\right|
- \frac{p}{2}\sum_{g=1}^G n_{g,\text{bad}}^{(r)} \ln\rho_g
- \frac{p}{2}\sum_{i=1}^n\sum_{g=1}^G z_{ig}^{(r)} \left(1-v_{ig}^{(r)}\right) \widetilde{E}_{3ig}^{(r)}
\nonumber\\
&
- \frac{1}{2}\sum_{i=1}^n\sum_{g=1}^G z_{ig}^{(r)}\left(1-v_{ig}^{(r)}\right) \widetilde{E}_{2ig}^{(r)} \frac{1}{\rho_g}\left(\bx_i-\bmu_g\right)^{\top}\bSigma_g^{-1} \left(\bx_i-\bmu_g\right) 
\nonumber\\
&
+ \sum_{i=1}^n\sum_{g=1}^G z_{ig}^{(r)}\left(1-v_{ig}^{(r)}\right) \frac{1}{\sqrt{\rho_g}} \left(\bx_i-\bmu_g\right)^{\top} \bSigma_g^{-1} \balpha_g
\nonumber\\
&
- \frac{1}{2}\sum_{i=1}^n\sum_{g=1}^G z_{ig}^{(r)}\left(1-v_{ig}^{(r)}\right) \widetilde{E}_{1ig}^{(r)} \sqrt{\rho_g}\balpha_g^{\top}\bSigma_g^{-1}\balpha_g
- \sum_{i=1}^n\sum_{g=1}^G z_{ig}^{(r)}\left(1-v_{ig}^{(r)}\right) \widetilde{E}_{1ig}^{(r)}, 
\label{eq:expected complete-data log-likelihood - bad}
\end{align}
with $n_g^{(r)}=\sum_{i=1}^n z_{ig}^{(r)}$ being the expected size of group $g$, $n_{g,\text{good}}^{(r)}=\sum_{i=1}^n z_{ig}^{(r)}v_{ig}^{(r)}$ the expected number of good observations in group $g$, and $n_{g,\text{bad}}^{(r)}=\sum_{i=1}^n z_{ig}^{(r)}(1-v_{ig}^{(r)})$ the expected number of bad observations in group $g$.
Inside the formulae \eqref{eq:expected complete-data log-likelihood - pi}--\eqref{eq:expected complete-data log-likelihood - bad} there are the following updates 
\begin{equation*}
z_{ig}^{(r)} \colonequals E\left(Z_{ig}=1|\bx_i,\bvartheta^{(r)}\right) = 
\frac{\pi_g^{(r)} f_{\text{CSAL}}\left(\bx_i | \bmu_g^{(r)}, \bSigma_g^{(r)}, \balpha_g^{(r)}, \lambda_g^{(r)}, \rho_g^{(r)} \right)}{\displaystyle\sum_{h=1}^G\pi_h f_{\text{CSAL}}\left(\bx_i | \bmu_h^{(r)}, \bSigma_h^{(r)}, \balpha_h^{(r)}, \lambda_h^{(r)}, \rho_h^{(r)} \right)},
%\label{eq:update z}
\end{equation*}
\begin{equation*}
v_{ig}^{(r)} \colonequals E\left(V_{ig}=1|\bx_i,Z_{ig}=1,\bvartheta^{(r)}\right) = \frac{\lambda_g^{(r)} f_{\text{SAL}}\left(\bx_i | \bmu_g^{(r)}, \bSigma_g^{(r)}, \balpha_g^{(r)}\right)}{f_{\text{CSAL}}\left(\bx_i | \bmu_g^{(r)}, \bSigma_g^{(r)}, \balpha_g^{(r)}, \lambda_g^{(r)}, \rho_g^{(r)}\right)},
%\label{eq:salcontv}
\end{equation*} 
\begin{equation}
E_{1ig}^{(r)}\colonequals E\left(W_{ig} | \bx_i, Z_{ig}=1,\bvartheta^{(r)}\right) = \frac{\sqrt{b_{ig}^{(r)}}K_{\nu+1}\left(\sqrt{a_g^{(r)}b_{ig}^{(r)}}\right)}{\sqrt{a_g^{(r)}}K_{\nu}\left(\sqrt{a_g^{(r)}b_{ig}^{(r)}}\right)} ,   %\sqrt{\frac{a_{ig}}{b_g}}K_{\nu}\left(u\right),
\label{eq:E1ig good}
\end{equation}
\begin{equation}
E_{2ig}^{(r)}\colonequals E\left(W_{ig}^{-1} | \bx_i, Z_{ig}=1,\bvartheta^{(r)}\right) = \frac{\sqrt{a_g^{(r)}}K_{\nu+1}\left(\sqrt{a_g^{(r)}b_{ig}^{(r)}}\right)}{\sqrt{b_{ig}^{(r)}}K_{\nu}\left(\sqrt{a_g^{(r)}b_{ig}^{(r)}}\right)} - \frac{2\nu}{b_{ig}^{(r)}}, %\sqrt{\frac{b_g}{a_{ig}}}K_{\nu}\left(u\right)-\frac{2\nu}{a_{ig}},
\label{eq:E2ig good}
\end{equation}
\begin{equation}
E_{3ig}^{(r)}\colonequals E\left(\ln W_{ig} | \bx_i, Z_{ig}=1,\bvartheta^{(r)}\right) = 
\ln \frac{\sqrt{b_{ig}^{(r)}}}{\sqrt{a_g^{(r)}}} + \frac{\partial}{\partial \nu}\ln K_{\nu}\left(\sqrt{a_g^{(r)}b_{ig}^{(r)}}\right),
\label{eq:E3ig good}
\end{equation}
\begin{equation}
\widetilde{E}_{1ig}^{(r)} \colonequals E\left(\widetilde{W}_{ig} | \bx_i, Z_{ig}=1,\bvartheta^{(r)}\right) = \frac{\sqrt{\widetilde{b}_{ig}^{(r)}}K_{\nu+1}\left(\sqrt{a_g^{(r)}\widetilde{b}_{ig}^{(r)}}\right)}{\sqrt{a_g^{(r)}}K_{\nu}\left(\sqrt{a_g^{(r)}\widetilde{b}_{ig}^{(r)}}\right)},
\label{eq:E1ig bad}
\end{equation}
\begin{equation}
\widetilde{E}_{2ig}^{(r)} \colonequals E\left(\widetilde{W}_{ig}^{-1} | \bx_i, Z_{ig}=1,\bvartheta^{(r)}\right) = \frac{\sqrt{a_g^{(r)}}K_{\nu+1}\left(\sqrt{a_g^{(r)}\widetilde{b}_{ig}^{(r)}}\right)}{\sqrt{\widetilde{b}_{ig}^{(r)}}K_{\nu}\left(\sqrt{a_g^{(r)}\widetilde{b}_{ig}^{(r)}}\right)} - \frac{2\nu}{\widetilde{b}_{ig}^{(r)}},
\label{eq:E2ig bad}
\end{equation}
and
\begin{equation}
\widetilde{E}_{3ig}^{(r)}\colonequals E\left(\ln \widetilde{W}_{ig} | \bx_i, Z_{ig}=1,\bvartheta^{(r)}\right) =
\ln \frac{\sqrt{\widetilde{b}_{ig}^{(r)}}}{\sqrt{a_g^{(r)}}} + \frac{\partial}{\partial \nu}\ln K_{\nu}\left(\sqrt{a_g^{(r)}\widetilde{b}_{ig}^{(r)}}\right) 
\label{eq:E3ig bad}
\end{equation}
where $a_g^{(r)}=2+(\balpha_g^{(r)})^{\top}(\bSigma_g^{(r)})^{-1}\balpha_g^{(r)}$, $b_{ig}^{(r)}=\delta(\bx_i,\bmu_g^{(r)}|\bSigma_g^{(r)})$, $\widetilde{b}_{ig}^{(r)}=\delta(\bx_i,\bmu_g^{(r)}|\rho_g^{(r)}\bSigma_g^{(r)})$ and $\nu = (2 - p)/2$.
The closed forms for $E_{1ig}^{(r)}$, $E_{2ig}^{(r)}$, $E_{3ig}^{(r)}$, $\widetilde{E}_{1ig}^{(r)}$, $\widetilde{E}_{2ig}^{(r)}$ and $\widetilde{E}_{3ig}^{(r)}$ in \eqref{eq:E1ig good}--\eqref{eq:E3ig bad} exist because $W_{ig}|\bx_1,Z_{ig}=1\sim \mathcal{GIG}(a_g,b_{ig},(2 - p)/2)$ and $\widetilde{W}_{ig}|\bx_1,Z_{ig}=1\sim \mathcal{GIG}(a_g,\widetilde{b}_{ig},(2 - p)/2)$, where $\mathcal{GIG}\left(a,b,\nu\right)$ denotes the generalized inverse Gaussian (GIG) distribution with parameters $a>0$, $b>0$ and $\nu\in\real$.  
Note that the terms of the Q-function where $E_{3ig}^{(r)}$ and $\widetilde{E}_{3ig}^{(r)}$ appear --- refer to \eqref{eq:expected complete-data log-likelihood - good} and \eqref{eq:expected complete-data log-likelihood - bad} --- are constant with respect to the model parameters $\bvartheta$; therefore, $E_{3ig}^{(r)}$ and $\widetilde{E}_{3ig}^{(r)}$ are not required in our calculations but we provide them for completeness. 

%%%%%%%%%%%%%
% CM-step 1 %
%%%%%%%%%%%%%

\subsection{CM-step 1}
\label{subsec:CM-step 1}

The first CM-step requires the calculation of $\bvartheta_1^{\left(r+1\right)}$ as the value of $\bvartheta_1$ that maximizes $Q(\bvartheta)$ with $\bvartheta_2$ fixed at $\bvartheta_2^{(r)}$.
In particular, the maximization of $Q_1(\bpi)$ with respect to $\bpi$, subject to the constraints on these parameters, yields
%after some algebra, we obtain
\begin{equation*}
\pi_g^{(r+1)} = \frac{n_g^{(r)}}{n}, \quad g=1,\ldots,G.
%\label{eq:pi update}
\end{equation*}
Analogously, the maximization of $Q_2(\blambda)$ with respect to the generic element $\lambda_g$ of $\blambda$, $g=1,\ldots,G$, leads to
\begin{equation*}
\lambda_g^{(r+1)} = \frac{n_{g,\text{good}}^{(r)}}{n_g^{(r)}}.
%\label{eq:lambda update}
\end{equation*}
Finally, the maximization of $Q_3^{\text{good}}(\bmu,\bSigma,\balpha)$ and 
$Q_3^{\text{bad}}(\bmu,\bSigma,\balpha,\brho)$, with respect to the $g$th elements of $\bmu$, $\bSigma$ and $\balpha$, $g=1,\ldots,G$, yields
\begin{equation*}
\bmu_g^{(r+1)}=\frac{B^{(r)}\left[\displaystyle\sum_{i=1}^n z_{ig}^{(r)} \left(v_{ig}^{(r)}E_{2ig}^{(r)}+\frac{1-v_{ig}^{(r)}}{\rho_g^{(r)}}\widetilde{E}_{2ig}^{(r)}\right)\bx_i\right]-C^{(r)}\left[\displaystyle\sum_{i=1}^nz_{ig}^{(r)}\left(v_{ig}^{(r)}+\frac{1-v_{ig}^{(r)}}{\rho_g^{(r)}}\right)\bx_i\right]}{B^{(r)}A^{(r)}-\left(C^{(r)}\right)^2},
%\label{eq:mu update}
\end{equation*}
\begin{equation*}
\balpha_g^{(r+1)}=\frac{A^{(r)}\left[\displaystyle\sum_{i=1}^nz_{ig}^{(r)}\left(v_{ig}^{(r)}+\frac{1-v_{ig}^{(r)}}{\sqrt{\rho_g^{(r)}}}\right)\bx_i\right]-D^{(r)}\left[\displaystyle\sum_{i=1}^n z_{ig}^{(r)} \left(v_{ig}^{(r)}E_{2ig}^{(r)}+\frac{1-v_{ig}^{(r)}}{\rho_g^{(r)}}\widetilde{E}_{2ig}^{(r)}\right)\bx_i\right]} {B^{(r)}A^{(r)}-\left(D^{(r)}\right)^2},
%\label{eq:alpha update}
\end{equation*}
and 
\begin{align*}
\bSigma_g^{(r+1)}=
&
\frac{1}{n_g^{(r)}}\sum_{i=1}^n z_{ig}^{(r)} \left(v_{ig}^{(r)}E_{2ig}^{(r)}+\frac{1-v_{ig}^{(r)}}{\rho_g^{(r)}}\widetilde{E}_{2ig}^{(r)}\right)\left(\bx_i-\bmu_g^{(r+1)}\right)\left(\bx_i-\bmu_g^{(r+1)}\right)^{\top}
\\
&
- \frac{2}{n_g^{(r)}}\sum_{i=1}^nz_{ig}^{(r)}\left(v_{ig}^{(r)}+\frac{1-v_{ig}^{(r)}}{\sqrt{\rho_g^{(r)}}} \right) \left(\bx_i-\bmu_g^{(r+1)}\right)\left(\balpha_g^{(r+1)}\right)^{\top}
\\
%&
%-\frac{1}{n_g^{(r)}}\sum_{i=1}^nz_{ig}^{(r)}\left(v_{ig}^{(r)}+\frac{1-v_{ig}^{(r)}}{\sqrt{\rho_g^{(r)}}}\right) \balpha_g^{(r+1)} \left(\bx_i-\bmu_g^{(r+1)}\right)^{\top}
%\nonumber\\
&
+\frac{1}{n_g^{(r)}}\sum_{i=1}^n z_{ig}^{(r)} \left[v_{ig}^{(r)}E_{1ig}^{(r)}+\left(1-v_{ig}^{(r)}\right)\widetilde{E}_{1ig}^{(r)}\right] \balpha_g^{(r+1)}\left(\balpha_g^{(r+1)}\right)^{\top},
%\label{eq:Sigma update}
\end{align*}
where
\begin{equation*}
A^{(r)}=\sum_{i=1}^n z_{ig}^{(r)} \left(v_{ig}^{(r)}E_{2ig}^{(r)}+\frac{1-v_{ig}^{(r)}}{\rho_g^{(r)}}\widetilde{E}_{2ig}^{(r)}\right),
%\label{eq:A}
\end{equation*}
\begin{equation*}
B^{(r)}=\sum_{i=1}^n z_{ig}^{(r)} \left[v_{ig}^{(r)}E_{1ig}^{(r)}+\left(1-v_{ig}^{(r)}\right)\widetilde{E}_{1ig}^{(r)}\right],
%\label{eq:B}
\end{equation*}
\begin{equation*}
C^{(r)}=\sum_{i=1}^n z_{ig}^{(r)} \left(v_{ig}^{(r)}+\frac{1-v_{ig}^{(r)}}{\rho_g^{(r)}}\right),
%\label{eq:C}
\end{equation*}
and
\begin{equation*}
D^{(r)}=\sum_{i=1}^n z_{ig}^{(r)} \left(v_{ig}^{(r)}+\frac{1-v_{ig}^{(r)}}{\sqrt{\rho_g^{(r)}}}\right).
%\label{eq:D}
\end{equation*}

%%%%%%%%%%%%%
% CM-step 2 %
%%%%%%%%%%%%%

\subsection{CM-step 2}
\label{subsec:CM-step 2}

The second CM-step requires the calculation of $\bvartheta_2^{\left(r+1\right)}$ as the value of $\bvartheta_2$ that maximizes $Q(\bvartheta)$ with $\bvartheta_1$ fixed at $\bvartheta_1^{\left(r+1\right)}$.
In particular, we have to maximize $Q_4^{\text{bad}}\left(\bmu=\bmu^{(r+1)},\bSigma=\bSigma^{(r+1)},\balpha=\balpha^{(r+1)},\brho\right)$ with respect to each element $\rho_g$ of $\brho$, $g = 1, \ldots ,G$, under the constraint $\rho_g>1$.
%In \eqref{eq:rho update}, terms independent of $\rho_g$ are omitted.
Operationally, the \texttt{optimize()} function, in the \texttt{stats} package for \textsf{R}, can be used to perform a numerical search of the maximum.

\section{Further computational and operational aspects}
\label{sec:Further aspects}

\subsection{\textcolor{black}{Initialization strategy}}
\label{subsec:Initialization strategy}

The choice of the initialization strategy for EM-based algorithms constitutes an important issue (see, e.g., \citealp{Bier:Cele:Gova:Choo:2003} and \citealp{Karl:Xeka:Choo:2003}).
For the ECM algorithm described in Section~\ref{sec:Parameter estimation}, we adopt an initialization strategy arising by the fact that the $G$-component SAL mixture can be seen as a special case of the $G$-component CSAL mixture when $\lambda_g\rightarrow 1$ and $\rho_g\rightarrow 1$, $g=1,\ldots,G$ (nested models).
Specifically, we initialize the EM algorithm to fit the $G$-component SAL mixture \citep[see][for details]{franczak14} by providing the initial indicator vectors $\bz_1^{(0)},\ldots,\bz_n^{(0)}$ for the first M-step; we obtain these vectors by a preliminary run of the $k$-means method (as implemented by the \texttt{kmeans()} function of the \textbf{stats} package for \textsf{R}).
The obtained estimates of $\pi_1,\ldots,\pi_G$, $\bmu_1,\ldots,\bmu_G$, $\bSigma_1,\ldots,\bSigma_G$, and $\balpha_1,\ldots,\balpha_G$ for the SAL mixture, along with the values $\lambda_1=\cdots=\lambda_G=0.999$ and $\rho_1=\cdots=\rho_G=1.001$, are used to initialize the first E-step of the ECM algorithm to fit the $G$-component CSAL mixture.  
From an operational point of view, such an initialization strategy, thanks to the monotonicity property of the ECM algorithm \citep[see, e.g.,][p.~28]{McLa:Kris:TheE:2007}, guarantees that the observed-data log-likelihood of the CSAL mixture will be always greater than, or equal to, the observed-data log-likelihood of the SAL mixture.
See \citet{Punz:McNi:Robu:2016,Punz:McNi:Robu:2017} and \citet{Mazz:Punz:Mixt:2018} for an analogous initialization strategy applied to mixtures based on the contaminated normal distribution.

\subsection{Convergence criterion}
\label{subsec:Convergence criterion}

We use a stopping criterion based on the Aitken acceleration \citep{Aitk:OnBe:1926} to determine convergence of the ECM algorithm illustrated in Section~\ref{sec:Parameter estimation}. 
The commonly used stopping rules can yield convergence earlier than the Aitken stopping criterion, resulting in estimates that might not be close to the ML estimates. 
The Aitken acceleration at iteration $r$ is
\begin{equation*}
a^{\left(r\right)}=\frac{l^{\text{new}}-l^{\left(r\right)}}{l^{\left(r\right)}-l^{\left(r-1\right)}},
\end{equation*}
where $l^{\left(r\right)}$ is the (observed-data) log-likelihood from iteration $r$. 
An asymptotic --- with respect to the iteration number --- estimate of the log-likelihood at iteration $r + 1$ can be computed via
\begin{equation*}	
l_A^{\text{new}}=l^{\left(q\right)}+\frac{1}{1-a^{\left(r\right)}}\left(l^{\text{new}}-l^{\left(r\right)}\right),
\end{equation*}
see~\citet{Bohn:Diet:Scha:Schl:Lind:TheD:1994}. 
Convergence is assumed to have been reached when $l_A^{\text{new}}-l^{\left(r\right)}<\epsilon$, provided that this difference is positive (see \citealp{McNi:Murp:McDa:Fros:Seri:2010}). 
We use $\epsilon = 10^{-15}$ in the analyses of Section~\ref{sec:Applications to artificial and real data}.

%\clearpage
%
%After convergence, component memberships are usually estimated based on the maximum {\it a posteriori} (MAP) classification given by $\text{MAP}\{{\hat z}_{ig}\}=1$ if $\text{max}_h\{{\hat z}_{ih}\}$ occurs at component $g$, and $\text{MAP}\{{\hat z}_{ig}\}=0$ otherwise.
%%
%Note that the MAP classification is used to give the predicted classifications (clusterings) in the data analyses presented herein.

\subsection{\textcolor{black}{Dealing with infinite log-likelihood values}}
\label{subsec:Dealing with infinite log-likelihood values}

As well-documented in \citet[][Section~3.4.2]{franczak14} in the case of the SAL mixture, complications may arise when updating the component mean vectors if $\bmu_g^{(r)}$, $g=1,\ldots,G$, tends to an observation $\bx_i$.
In such situations, computational issues arise when we try to determine the remaining parameter values and, specifically, the expected values $E_{2ig}^{(r)}$ and $\widetilde{E}_{2ig}^{(r)}$.   
To overcome this problem, we 
%follow \citet[][Section~3.4.2]{franczak14} and 
stop updating $\bmu_g$ when its value equals some $\bx_i$. 
Details about how to implement this simple-minded procedure can be found in \citet[][Section~3.4.2]{franczak14}.

\subsection{\textcolor{black}{Model selection}}
\label{subsec:BIC}

In model-based clustering applications, it is common to fit a mixture model for a range of values of the number of components $G$.
After that, the ``best'' value for $G$ is chosen based on some likelihood-based criterion, although such a choice does not necessarily correspond to optimal clustering; for the alternative use of likelihood-ratio tests, see \citet{Punz:Brow:McNi:Hypo:2016}. 

The Bayesian information criterion \citep[BIC;][]{Schw:Esti:1978} is commonly used for mixture model selection. 
It is intended to provide a measure of the weight of evidence favoring one model over another, or Bayes factor \citep{Weak:Acri:1999}.
Even though the regularity properties needed for the development of the BIC are not satisfied by mixture models \citep{Keri:Cons:2000}, it has been used extensively (see, e.g., \citealp{Dasg:Raft:Dete:1998} and \citealp{Fral:Raft:Mode:2002}) and performs well in practice. 
The BIC can be computed as
\begin{equation*}
\text{BIC} = 2l(\hat{\boldsymbol{\vartheta}}) - \kappa\ln n,
%\label{eq:BIC}
\end{equation*}
where $l(\hat{\boldsymbol{\vartheta}})$ is the maximized (observed) log-likelihood, $\kappa$ is the number of free parameters, and $n$ is the sample size.
Note that, Bayes factors can be used to compare models that are not nested, and the BIC approximation thereto holds under equal priors (see~\citealp{Raft:Baye:1995,Dasg:Raft:Dete:1998}). 

\subsection{Performance assessment}
\label{subsec:Performance assessment}

\textcolor{black}{The adjusted Rand index \citep[ARI;][]{Hube:Arab:Comp:1985} is one of the methods we use --- in addition to the analysis of the confusion matrix and to the misclassification error rate --- for determining the classification performance of the chosen model by comparing predicted classifications to true group labels, when known; labels can be known, for example, when data are simulated with a known group-structure (see the analyses in Section~\ref{subsec:Assessing the impact of background noise}).}
The ARI corrects the classical Rand index \citep{Rand:Obje:1971} to account for chance agreement when comparing true labels and estimated classifications. 
An ARI of 1 corresponds to perfect agreement, and the expected value of the ARI is 0 under random classification.
Negative ARI values are possible and indicate classification results that are worse, in some sense, than would be expected by random classification.
\citet{Stei:Prop:2004} provides a thorough evaluation of the ARI.

\subsection{Automatic detection of bad points}
\label{subsec:Automatic detection of bad points}

For a contaminated mixture, the classification of an observation $\bx_i$ means: 
\begin{description}
	\item[Step 1.] Determine its cluster membership.
	\item[Step 2.] Establish whether it is a good or a bad observation in that cluster.
\end{description}
Let $\hat{\boldsymbol{z}}_i$ and $\hat{\boldsymbol{v}}_i$ denote, respectively, the expected values of $\boldsymbol{z}_i$ and $\boldsymbol{v}_i$ arising from the ECM algorithm, i.e., $\hat{z}_{ig}$ is the value of $z_{ig}^{\left(r\right)}$ at convergence and $\hat{v}_{ig}$ is the value of $v_{ig}^{\left(r\right)}$ at convergence. 
To evaluate the cluster membership of $\bx_i$, we use the maximum \textit{a~posteriori} (MAP) classification, i.e., $\text{MAP}\left(\hat{z}_{ig}\right)=1$ if $\arg\max_h\{\hat{z}_{ih}\}=g$ and $\text{MAP}\left(\hat{z}_{ig}\right)=0$ otherwise.
We then consider $\hat{v}_{ih}$, where $h$ is selected such that $\text{MAP}\left(\hat{z}_{ih}\right)=1$, and $\bx_i$ is considered good if $\hat{v}_{ih}>0.5$ and $\bx_i$ is considered bad otherwise. 
The resulting information can be used to eliminate the bad points, if such an outcome is desired \citep{Berk:Bent:Esti:1988}; the remaining data may then be treated as effectively being distributed according to a SAL mixture, and the clustering results can be reported as usual, i.e., based on $\text{MAP}\left(\hat{z}_{ig}\right)$.

\section{\textcolor{black}{Applications to artificial and real data}}
\label{sec:Applications to artificial and real data}

In this section, we evaluate the effectiveness and further aspects of our model through artificial and real data.
Other well-known (mixture) models are considered for comparison.
The whole analysis is conducted in \textsf{R} \citep{R:2017}.

\subsection{Sensitivity study}
\label{subsec: Sensitivity analysis}

A sensitivity study is here described with a twofold aim.
Firstly, we want to compare how outliers affect the ML estimates of the (common) parameters $\bmu$, $\bSigma$ and $\balpha$ for the SAL and CSAL distributions.
Secondly, we are interested in evaluating if the location of the outliers in the direction determined by $\balpha$ can be less harmful.
With these aims, one hundred bivariate data sets ($p=2$), of size $n=100$, are randomly generated by a SAL distribution with parameters 
\begin{equation}
\bmu=\begin{bmatrix*}[r]
    0\\  
		0 \\
\end{bmatrix*}, 
\quad 
\bSigma=\begin{bmatrix*}[r]
 1 & 0 \\
 0 & 1 \\
\end{bmatrix*} 
\quad \text{and} \quad 
\balpha=\begin{bmatrix*}[r]
    0 \\  
		5 \\
\end{bmatrix*}.
\label{eq:generating parameters 2}
\end{equation}
Data are generated via the \texttt{raml()} function of the \textbf{LaplacesDemon} package \citep{LaplacesDemon}.
\figurename~\ref{fig:bankruptcy scatter} shows an example of scatter plot for one of the simulated data sets.
By looking at this plot, it is easy to realize that the skewness vector $\balpha$ in \eqref{eq:generating parameters 2} directs the scatter along the line $X_1=0$, which we can consider ``north'' for simplicity.
\begin{figure}[!ht]
\centering
\resizebox{0.4\textwidth}{!}{
\includegraphics{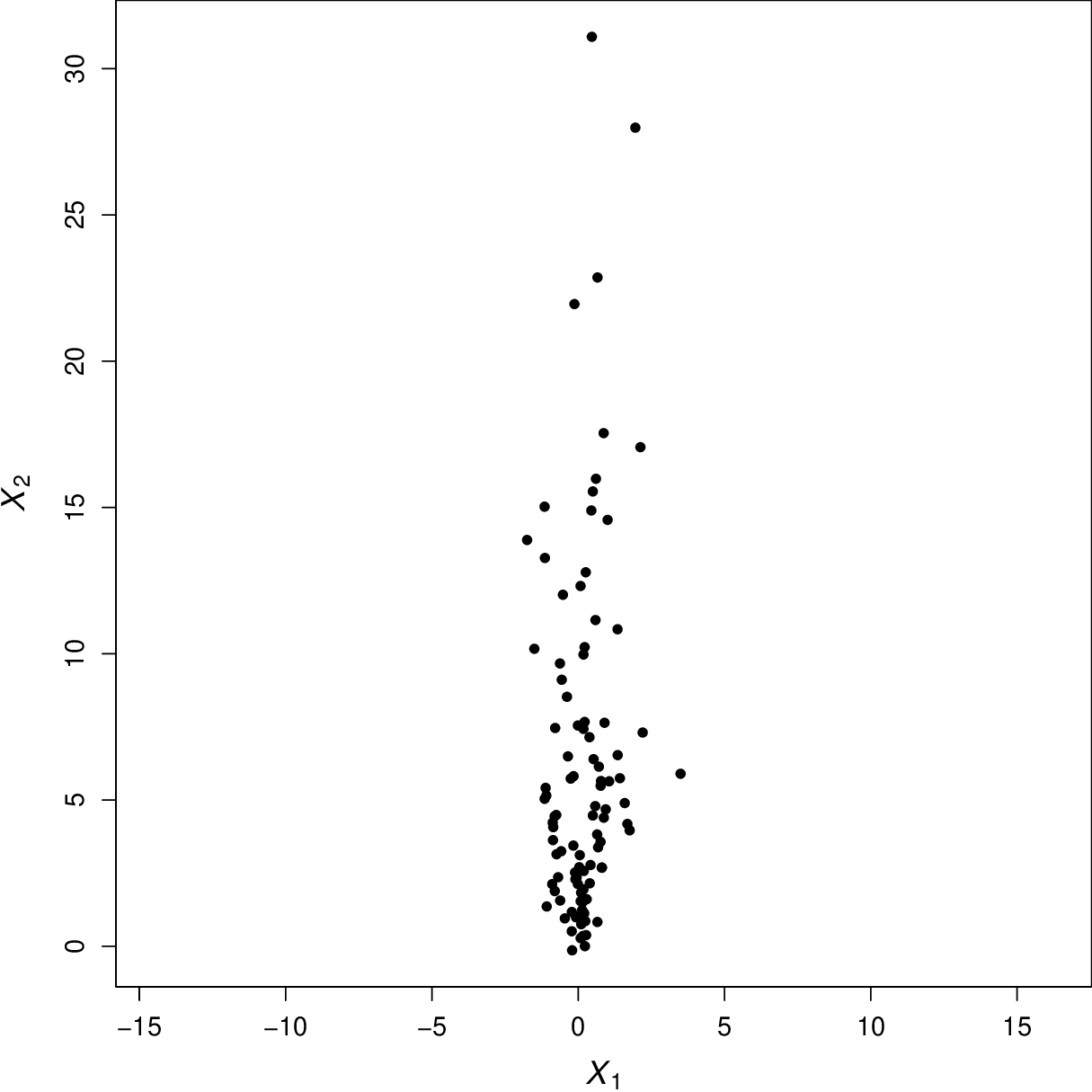} 
}
\caption{
Sensitivity analysis: example of generated data.
}
\label{fig:SAL data scatter}
\end{figure}

Nine perturbed versions of each generated data set are obtained by adding a single outlier in three of the four cardinal directions: outlier of coordinates $\left(0,y\right)^{\top}$ for the north, $\left(y,0\right)^{\top}$ for the east, and $\left(0,-y\right)^{\top}$ for the south, with $x>0$.
The west is not considered for its symmetry with the east.
Three levels of magnitude for the outliers are considered: small ($y=50$), medium ($y=75$) and large ($y=100$).
This results in a total of 900 perturbed data sets.
For each of them we estimate, by ML, the parameters of the SAL and CSAL distributions, using a convenient code implemented in \textsf{R} which is available upon request.

Each combination between the cardinal direction of the outlier (north, east, south) and its magnitude (small, medium, large), defines a different scenario.
Each scenario is characterized by one-hundred simulated data sets.
For each scenario, \tablename s~\ref{tab:Outlier on the top}--\ref{tab:Outlier on the bottom} report the average estimates (over the 100 simulated data sets) of the parameters for the SAL and CSAL distributions.
Regardless of the considered scenario, we have the following main findings.
\begin{itemize}
	\item ML estimation of $\bmu$, $\bSigma$ and $\balpha$ from the CSAL model is more robust to the added outlier, in the sense that the estimates are, in average, closer to the true parameters in \eqref{eq:generating parameters 2}.
	\item The estimated value of $\rho$ for the CSAL model increases in line with the magnitude of the outlier, so justifying its interpretation as degree of contamination, i.e.,~as a measure of how different outliers are from the bulk of the data; see also \citet{Punz:Maru:Clus:2016}.
	\item The outlier main affects the estimation of $\bSigma$ and $\balpha$, with the estimate of the latter being main biased in the variate the outlier acts.
	\item The outlier is certainly less harmful in the ``north'' scenario because it is closer to the scatter of the data (compare \tablename~\ref{tab:Outlier on the top} with \tablename s~\ref{tab:Outlier on the right}--\ref{tab:Outlier on the bottom}).
\end{itemize}
Summarizing, according to the obtained results, the CSAL distribution seems to be a good way to ``protect'' the SAL distribution in the presence of outliers, regardless of their location.
\renewcommand{\arraystretch}{1.1}
%%%%%%%%%
%% TOP %%
%%%%%%%%%
\begin{table}[!ht]
\caption{Outlier on the top: parameter estimates averaged over 100 replications.}
\label{tab:Outlier on the top}
\centering
%\resizebox{\textwidth}{!}{
\begin{tabular}{rlrrrrr}
\toprule
\multicolumn{1}{c}{Outlier} & \multicolumn{1}{c}{Model} & \multicolumn{1}{c}{$\bmu$} & \multicolumn{1}{c}{$\bSigma$} & \multicolumn{1}{c}{$\balpha$} & \multicolumn{1}{c}{$\lambda$} & \multicolumn{1}{c}{$\rho$} \\
\midrule
\multirow{2}{*}{
$
\begin{bmatrix*}[r]
    0\\  
		50 \\
\end{bmatrix*}
$
} 
& SAL & 
$\begin{bmatrix*}[r]
    -0.011\\  
		 0.001 \\
\end{bmatrix*}$
&
$\begin{bmatrix*}[r]
 1.065 & -0.043 \\
-0.043 &  1.085 \\
\end{bmatrix*}$
&
$\begin{bmatrix*}[r]
    0.008 \\ 
		5.401 \\
\end{bmatrix*}$
&
--
&
--
\\
& CSAL & 
$\begin{bmatrix*}[r]
    -0.011 \\
		 0.004 \\
\end{bmatrix*}$
&
$\begin{bmatrix*}[r]
1.013 & -0.024 \\
-0.024 & 0.973 \\
\end{bmatrix*}$
&
$\begin{bmatrix*}[r]
    0.019 \\
		5.159 \\
\end{bmatrix*}$
&
0.981
&
44.285
\\[6mm]
\multirow{2}{*}{
$\begin{bmatrix*}[r]
    0\\  
		75 \\
\end{bmatrix*}$
} & SAL & 
$\begin{bmatrix*}[r]
    -0.019 \\
		-0.002 \\
\end{bmatrix*}$
&
$\begin{bmatrix*}[r]
 1.103 & -0.064 \\
-0.064 & 1.145 \\
\end{bmatrix*}$
&
$\begin{bmatrix*}[r]
    0.016 \\
		5.626  \\
\end{bmatrix*}$
&
--
&
--
\\
& CSAL & 
$\begin{bmatrix*}[r]
    -0.010 \\  
		 0.004 \\
\end{bmatrix*}$
&
$\begin{bmatrix*}[r]
 1.008 & -0.020 \\
-0.020 &  0.951 \\
\end{bmatrix*}$
&
$\begin{bmatrix*}[r]
    0.012 \\
		5.112 \\
\end{bmatrix*}$
&
0.986
&
82.055
\\[6mm]
\multirow{2}{*}{
$\begin{bmatrix*}[r]
      0 \\  
		100 \\
\end{bmatrix*}$
} & SAL & 
$\begin{bmatrix*}[r]
    -0.014 \\  
		 0.002 \\
\end{bmatrix*}$
&
$\begin{bmatrix*}[r]
 1.127 & -0.059 \\
-0.059 &  1.342 \\
\end{bmatrix*}$
&
$\begin{bmatrix*}[r]
    0.012 \\ 
		5.823 \\
\end{bmatrix*}$
&
--
&
--
\\
& CSAL & 
$\begin{bmatrix*}[r]
    -0.012 \\  
		 0.005 \\
\end{bmatrix*}$
&
$\begin{bmatrix*}[r]
 1.015 & -0.021 \\
-0.021 &  0.944 \\
\end{bmatrix*}$
&
$\begin{bmatrix*}[r]
    0.014 \\
		5.122 \\
\end{bmatrix*}$
&
0.987
&
121.819
\\
\bottomrule
\end{tabular}
%}
\end{table}
%%%%%%%%%%%
%% RIGHT %%
%%%%%%%%%%%
\begin{table}[!ht]
\caption{Outlier on the right: parameter estimates averaged over 100 replications.}
\label{tab:Outlier on the right}
\centering
%\resizebox{\textwidth}{!}{
\begin{tabular}{rlrrrrr}
\toprule
\multicolumn{1}{c}{Outlier} & \multicolumn{1}{c}{Model} & \multicolumn{1}{c}{$\bmu$} & \multicolumn{1}{c}{$\bSigma$} & \multicolumn{1}{c}{$\balpha$} & \multicolumn{1}{c}{$\lambda$} & \multicolumn{1}{c}{$\rho$} \\
\midrule
\multirow{2}{*}{
$\begin{bmatrix*}[r]
    50\\  
		0 \\
\end{bmatrix*}$
} & SAL & 
$\begin{bmatrix*}[r]
    -0.029 \\
    -0.028 \\
\end{bmatrix*}$
&
$\begin{bmatrix*}[r]
 2.899 & -3.984 \\
-3.984 &  8.744 \\
\end{bmatrix*}$
&
$\begin{bmatrix*}[r]
    0.560 \\
    4.998 \\
\end{bmatrix*}$
&
--
&
--
\\
& CSAL & 
$\begin{bmatrix*}[r]
    -0.046 \\ 
    -0.033 \\
\end{bmatrix*}$
&
$\begin{bmatrix*}[r]
 1.061 &-0.499 \\
-0.499 & 1.311
\end{bmatrix*}$
&
$\begin{bmatrix*}[r]
    0.068 \\ 
    4.920 \\
\end{bmatrix*}$
&
0.981
&
2297.692
\\[6mm]
\multirow{2}{*}{
$\begin{bmatrix*}[r]
    75 \\  
		 0 \\
\end{bmatrix*}$
} & SAL & 
$\begin{bmatrix*}[r]
    -0.038 \\ 
    -0.026 \\
\end{bmatrix*}$
&
$\begin{bmatrix*}[r]
 3.810 & -5.514 \\
-5.514 & 11.410 \\
\end{bmatrix*}$
&
$\begin{bmatrix*}[r]
    0.797 \\ 
    5.003 \\
\end{bmatrix*}$
&
--
&
--
\\
& CSAL & 
$\begin{bmatrix*}[r]
    -0.039 \\ 
    -0.022 \\
\end{bmatrix*}$
&
$\begin{bmatrix*}[r]
 1.070 & -0.472 \\
-0.472 &  1.406 \\
\end{bmatrix*}$
&
$\begin{bmatrix*}[r]
    0.052\\
    4.885\\
\end{bmatrix*}$
&
0.981
&
9001.793
\\[6mm]
\multirow{2}{*}{
$\begin{bmatrix*}[r]
    100 \\  
		  0 \\
\end{bmatrix*}$
} & SAL & 
$\begin{bmatrix*}[r]
    -0.036\\ 
    -0.028\\
\end{bmatrix*}$
&
$\begin{bmatrix*}[r]
 4.684 &-6.802 \\
-6.802 &13.374 \\
\end{bmatrix*}$
&
$\begin{bmatrix*}[r]
    1.011\\
    4.947\\
\end{bmatrix*}$
&
--
&
--
\\
& CSAL & 
$\begin{bmatrix*}[r]
    -0.038 \\ 
    -0.018 \\
\end{bmatrix*}$
&
$\begin{bmatrix*}[r]
 1.126 &-0.487 \\
-0.487 & 1.400 \\
\end{bmatrix*}$
&
$\begin{bmatrix*}[r]
    0.052\\
    5.058\\
\end{bmatrix*}$
&
0.982
&
31184.950
\\
\bottomrule
\end{tabular}
%}
\end{table}
%%%%%%%%%%%%
%% BOTTOM %%
%%%%%%%%%%%%
\begin{table}[!ht]
\caption{Outlier on the bottom: parameter estimates averaged over 100 replications.}
\label{tab:Outlier on the bottom}
\centering
%\resizebox{\textwidth}{!}{
\begin{tabular}{rlrrrrr}
\toprule
\multicolumn{1}{c}{Outlier} & \multicolumn{1}{c}{Model} & \multicolumn{1}{c}{$\bmu$} & \multicolumn{1}{c}{$\bSigma$} & \multicolumn{1}{c}{$\balpha$} & \multicolumn{1}{c}{$\lambda$} & \multicolumn{1}{c}{$\rho$} \\
\midrule
\multirow{2}{*}{
$
\begin{bmatrix*}[r]
    0\\  
		-50 \\
\end{bmatrix*}
$
} 
& SAL & 
$\begin{bmatrix*}[r]
    0.085 \\
    0.074 \\
\end{bmatrix*}$
&
$\begin{bmatrix*}[r]
 0.944 & -0.165 \\
-0.165 & 22.389 \\
\end{bmatrix*}$
&
$\begin{bmatrix*}[r]
    -0.075 \\  
     4.038 \\
\end{bmatrix*}$
&
--
&
--
\\
& CSAL & 
$\begin{bmatrix*}[r]
    0.032 \\ 
    0.022 \\
\end{bmatrix*}$
&
$\begin{bmatrix*}[r]
 0.911 &-0.089 \\
-0.089 & 3.610 \\
\end{bmatrix*}$
&
$\begin{bmatrix*}[r]
    -0.026 \\  
     4.809 \\
\end{bmatrix*}$
&
    0.983
&
19725.990
\\[6mm]
\multirow{2}{*}{
$\begin{bmatrix*}[r]
    0\\  
		-75 \\
\end{bmatrix*}$
} & SAL & 
$\begin{bmatrix*}[r]
    0.082 \\ 
    0.079 \\
\end{bmatrix*}$
&
$\begin{bmatrix*}[r]
 0.999 & -0.366 \\
-0.366 & 31.140 \\
\end{bmatrix*}$
&
$\begin{bmatrix*}[r]
    -0.073 \\  
     3.755 \\
\end{bmatrix*}$
&
--
&
--
\\
& CSAL & 
$\begin{bmatrix*}[r]
    0.031 \\
    0.031 \\
\end{bmatrix*}$
&
$\begin{bmatrix*}[r]
 0.928 & -0.106 \\
-0.106 &  3.753 \\
\end{bmatrix*}$
&
$\begin{bmatrix*}[r]
    -0.026 \\
     4.827 \\
\end{bmatrix*}$
&
0.982
&
22695.340
\\[6mm]
\multirow{2}{*}{
$\begin{bmatrix*}[r]
      0 \\  
		-100 \\
\end{bmatrix*}$
} & SAL & 
$\begin{bmatrix*}[r]
    0.093 \\
    0.072 \\
\end{bmatrix*}$
&
$\begin{bmatrix*}[r]
 1.030 & -0.365 \\
-0.365 & 39.144 \\
\end{bmatrix*}$
&
$\begin{bmatrix*}[r]
    -0.072 \\
     3.517 \\
\end{bmatrix*}$
&
--
&
--
\\
& CSAL & 
$\begin{bmatrix*}[r]
    0.034 \\ 
    0.034 \\
\end{bmatrix*}$
&
$\begin{bmatrix*}[r]
 0.977 & -0.076 \\
-0.076 &  3.505 \\
\end{bmatrix*}$
&
$\begin{bmatrix*}[r]
    -0.030 \\
     4.916 \\
\end{bmatrix*}$
&
0.981
&
49268.330
\\
\bottomrule
\end{tabular}
%}
\end{table}

\subsection{Assessing the impact of background noise}
\label{subsec:Assessing the impact of background noise}

In this section we evaluate, through artificial data, how background noise can affect the classification performance of our mixture model.
Attention is also focused on the problem of detecting the underlying noise.
We further provide a comparison with finite mixtures of the following multivariate distributions:
%The main aim is to evaluate how background noise can affect the classification performance of the competing models.
%In this section we investigate the behaviour of the proposed mixture model through artificial data.
%We further provide a comparison with finite mixtures of some well-established multivariate distributions.
%The main aim is to evaluate how background noise can affect the classification performance of the competing models.
%Attention is also devoted to the problem of detecting the underlying noise.
%We compare mixtures of the following multivariate distributions: 
normal (N), $t$, contaminated normal (CN), skew-normal (SN), skew-$t$ (S$t$), skew-contaminated normal (SCN), skew-slash (SS), shifted asymmetric Laplace (SAL), and contaminated shifted asymmetric Laplace (CSAL).
To fit SAL and CSAL mixtures, we implemented a specific \textsf{R} code (available from the authors upon request).
CN mixtures are fitted via the \texttt{CNmixt()} function of the \textbf{ContaminatedMixt} package \citep{Punz:Mazz:McNi:ContaminatedMixt:2017,Punz:Mazz:McNi:Cont:2018}, while all the other models are fitted via the \texttt{smsn.mmix()} function of the \textbf{mixsmsn} package \citep{Prat:Lach:Cabr:mixsmsn:2013}.
Note that, for $t$, S$t$, SCN, and SS mixtures, the parameter(s) governing the tail-weight is (are) assumed to be equal across groups by the \textbf{mixsmsn} package. 
To allow for a direct comparison of the competing models, all the estimation algorithms are initialized by the solution provided by the $k$-means method (as implemented by the \texttt{kmeans()} function); see Section~\ref{subsec:Initialization strategy} with respect to the SAL and CSAL mixtures.

To simulate the ``good'' data of the following examples, we use a scheme not related to the models being fitted.
Specifically, following \citet{Franc:Thesis:2014}, we generate data in each group $g$, for $g=1,\ldots,G$, by a $p$-variate $t$ distribution with mean vector $\bmu_g$, scale matrix $\bSigma_g$ and $\nu_g$ degrees of freedom, but only the observations greater than $\bmu$ are retained, until the desired sample size is reached.
In the following, we refer to this procedure by saying that ``data are generated by a mixture of $p$-variate $t$-slice distributions''.
Moreover, more details are given in the first example of (i.e.,~Section~\ref{subsubsec: two groups and two dimensions}) while, for the sake of space, only essential insights are given in Sections~\ref{subsubsec: three groups and two dimensions}--\ref{subsubsec: two groups and three dimensions}.

\subsubsection{Example with $G=2$ groups and $p=2$ dimensions}
\label{subsubsec: two groups and two dimensions}

\begin{figure}[!ht]
\centering
\resizebox{0.4\textwidth}{!}{
\includegraphics{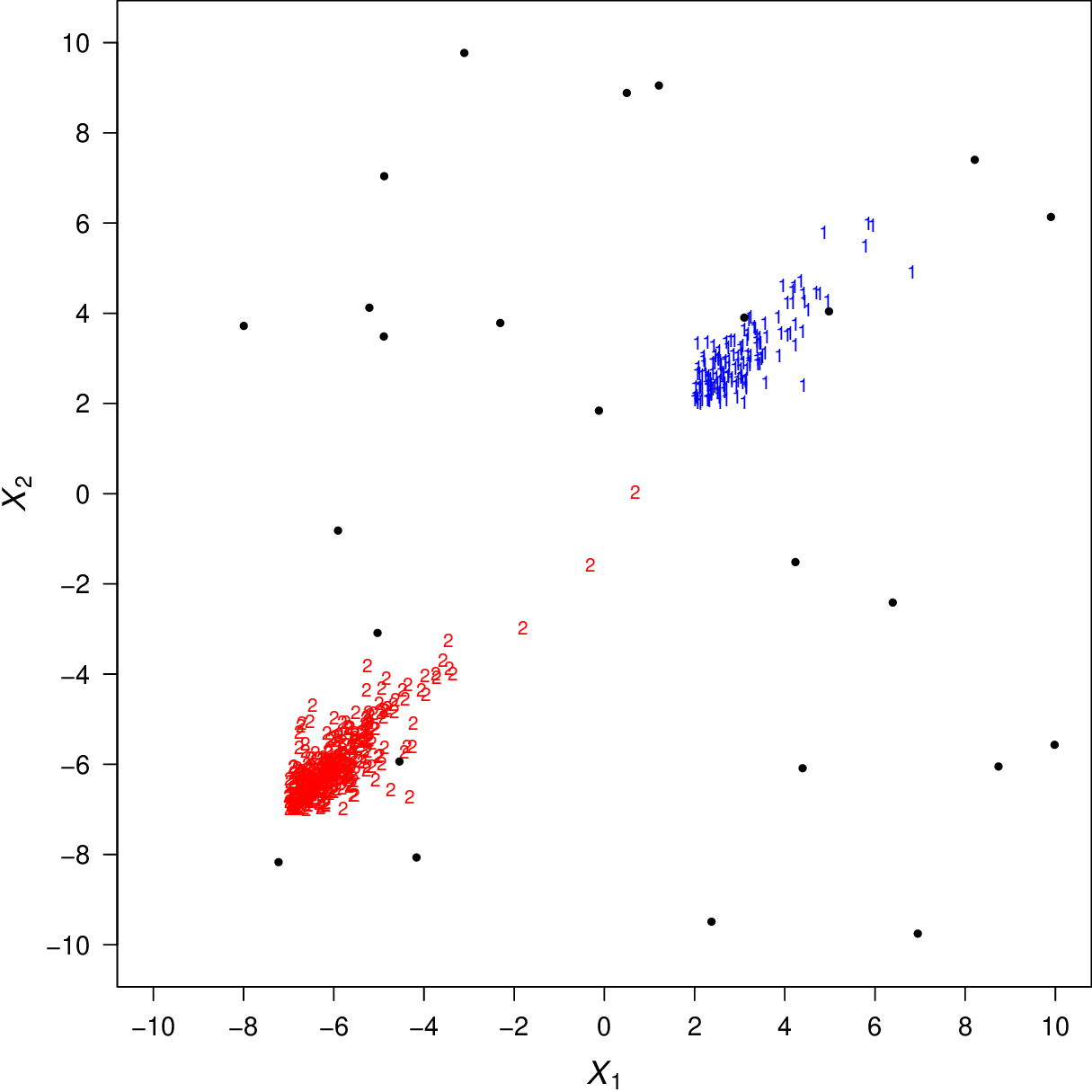} 
}
\caption{
Simulated data from Section~\ref{subsubsec: two groups and two dimensions}. 
Scatter plot (\textcolor{black}{1} and \textcolor{red}{2} denote the first and second group, respectively).
Background uniform noise points are denoted by $\bullet$.
}
\label{fig:noisescatter1}
\end{figure}

In this first example, $n=500$ bivariate ($p=2$) observations are generated by a mixture of $G=2$ $t$-slice distributions with parameters
\begin{equation*}
	\pi_1=0.3,
	\quad
	\bmu_1=
	\begin{bmatrix*}[c]
2	\\
2	 
\end{bmatrix*},
\quad 
	\bmu_2=
	\begin{bmatrix*}[c]
-7	\\
-7	 
\end{bmatrix*},
\quad \bSigma_1=\bSigma_2=\begin{bmatrix*}[c]
1 & 0.9	\\
0.9 & 1	 
\end{bmatrix*},
\quad \text{and} \quad 
	\nu_1=\nu_2=6.
%\label{eq:common generating parameters}
\end{equation*}
Moreover, 25 noise points are added from a uniform distribution over the interval $(-10, 10)$ on each dimension.
This yields an overall dataset comprising $n=525$ observations.
The scatter plot of the generated data, with labels \textcolor{black}{1} and \textcolor{red}{2} denoting points from the first and second group, respectively, and with bullets denoting uniform noise points, is displayed in \figurename~\ref{fig:noisescatter1}. 
Note that when a noise point effectively falls inside a cluster, which seems to happen at least three times (see \figurename~\ref{fig:noisescatter1}), we would expect it to be classified as belonging to the associated cluster.

\begin{figure}[!ht]
\centering
%%%%%%%%%%%%%%%
%% Symmetric %%
%%%%%%%%%%%%%%%
\subfigure[N mixture with $G=4$\label{fig:N.noisescatter1}]
{\resizebox{0.3\textwidth}{!}{\includegraphics{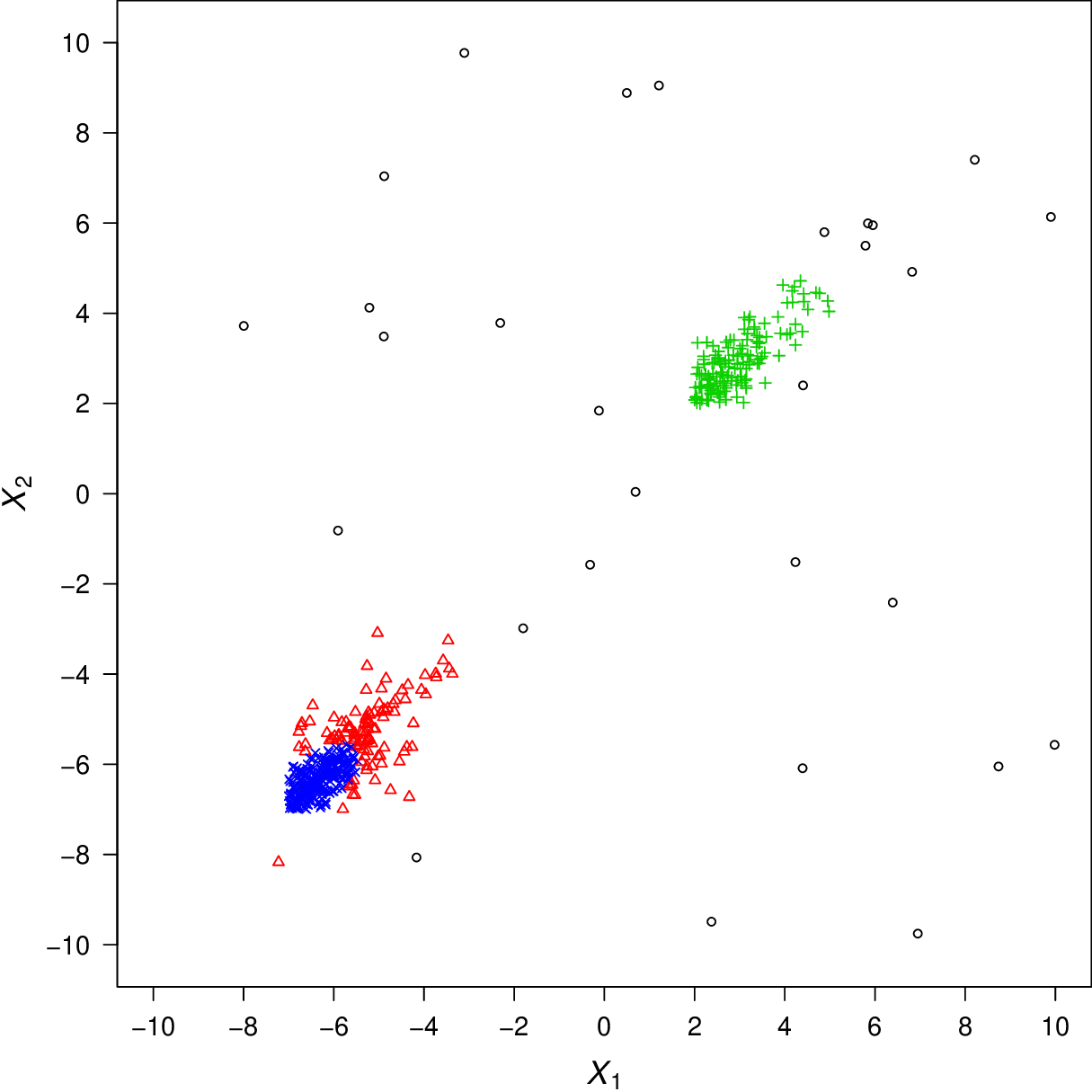}}}
%\quad
\subfigure[$t$ mixture with $G=4$\label{fig:t.noisescatter1}]
{\resizebox{0.3\textwidth}{!}{\includegraphics{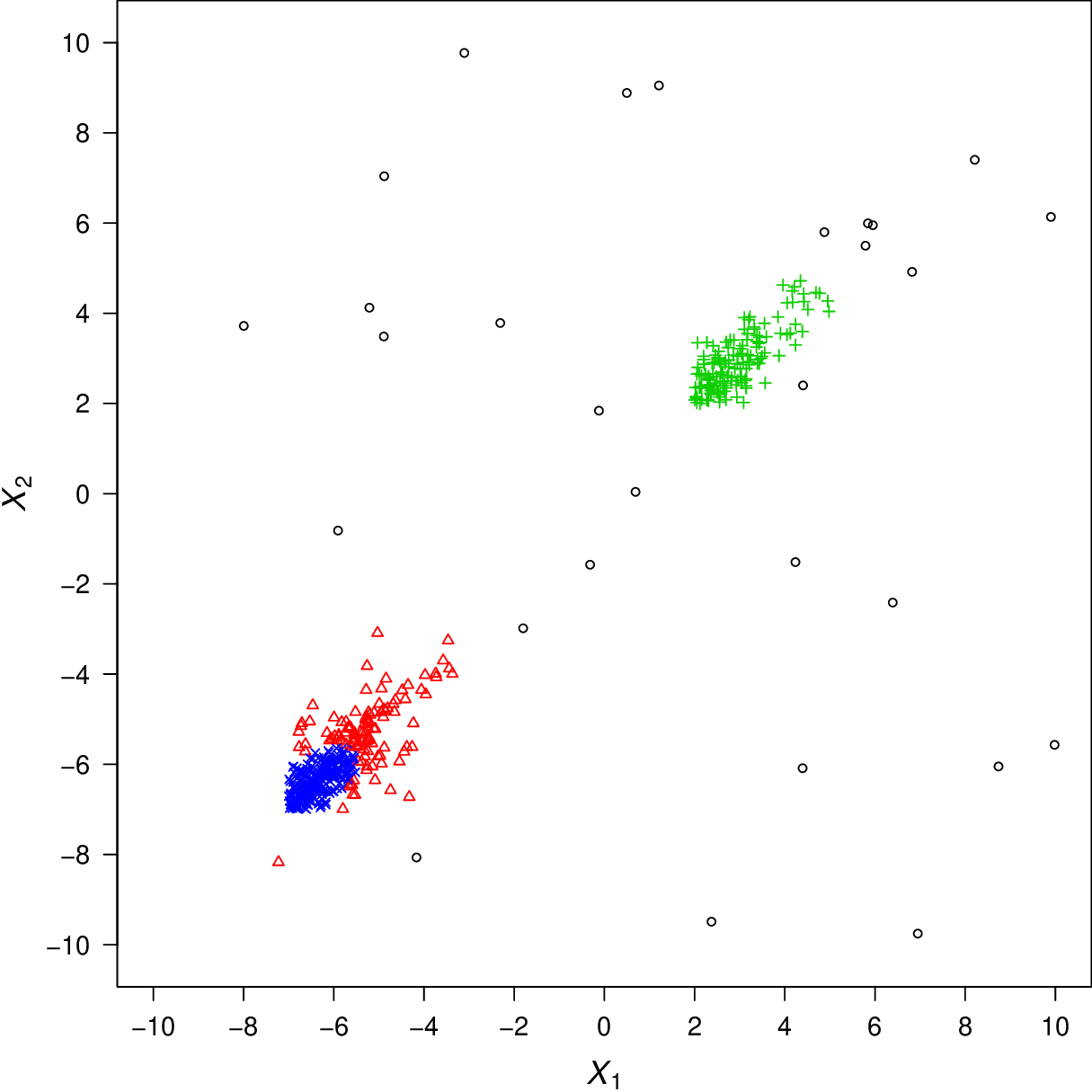}}}
%\quad
\subfigure[CN mixture with $G=4$\label{fig:CN.noisescatter1}]
{\resizebox{0.3\textwidth}{!}{\includegraphics{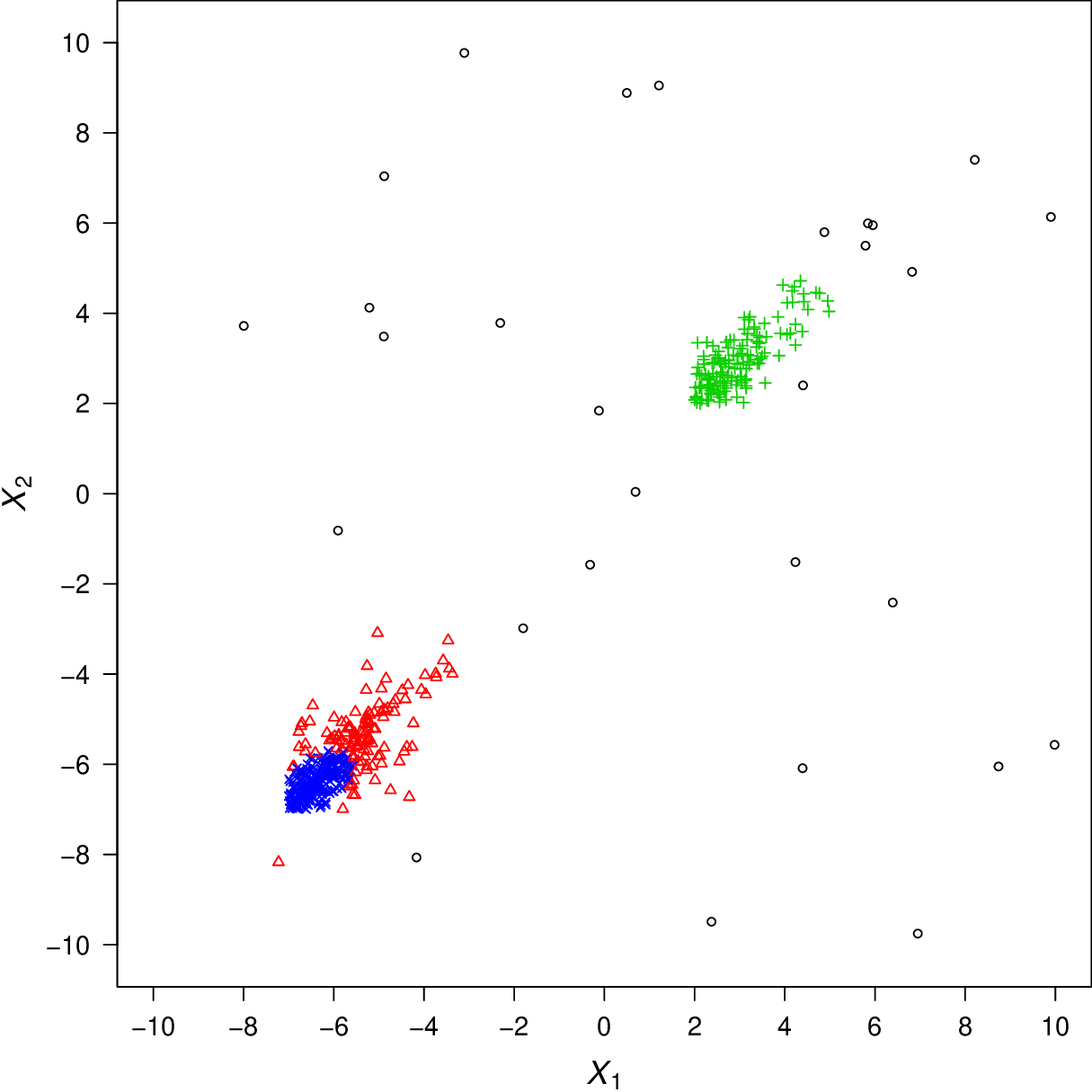}}}
%%%%%%%%%%%%%%%%
%% Asymmetric %%
%%%%%%%%%%%%%%%%
\subfigure[SN mixture with $G=4$\label{fig:SN.noisescatter1}]
{\resizebox{0.3\textwidth}{!}{\includegraphics{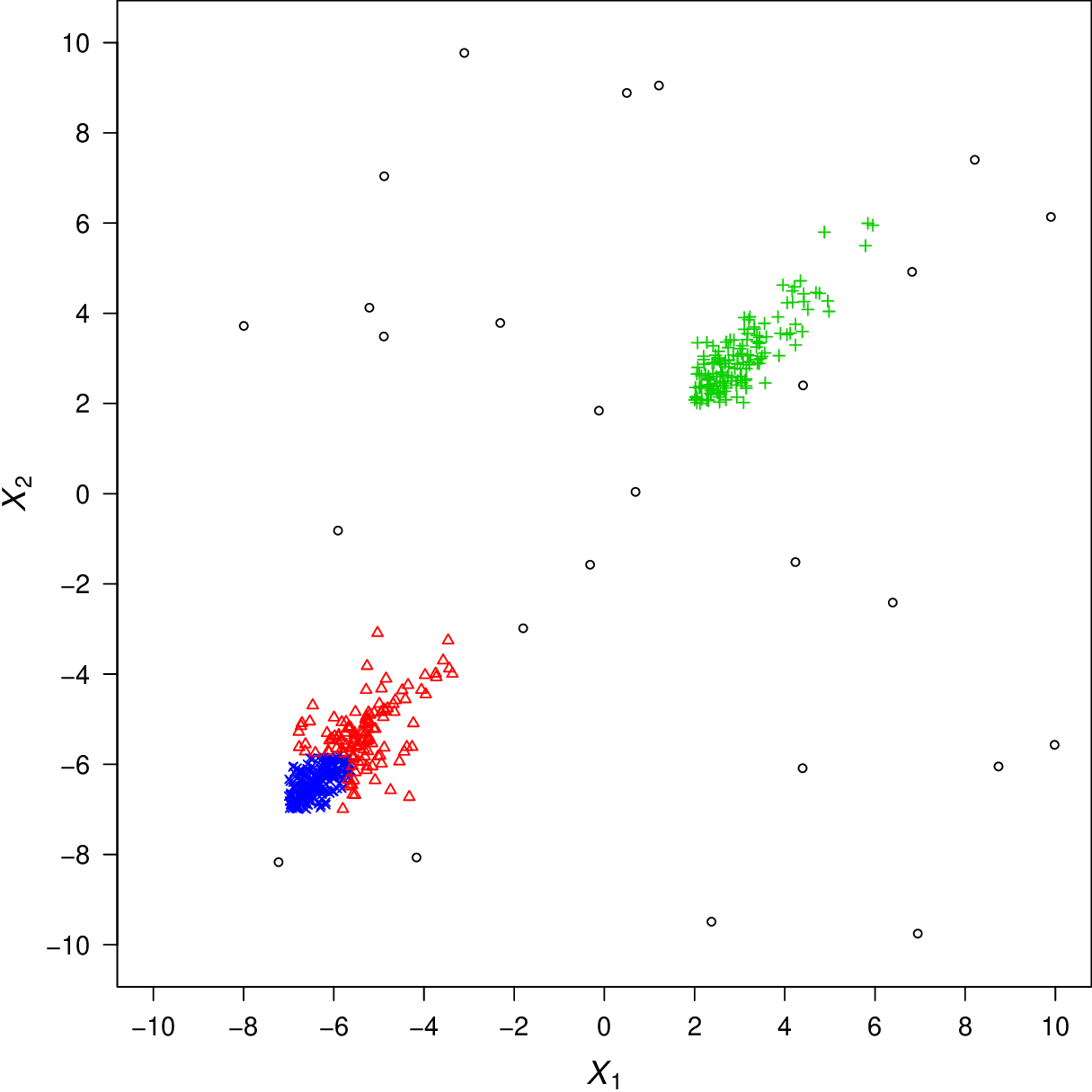}}}
%\quad
\subfigure[S$t$ mixture with $G=3$\label{fig:St.noisescatter1}]
{\resizebox{0.3\textwidth}{!}{\includegraphics{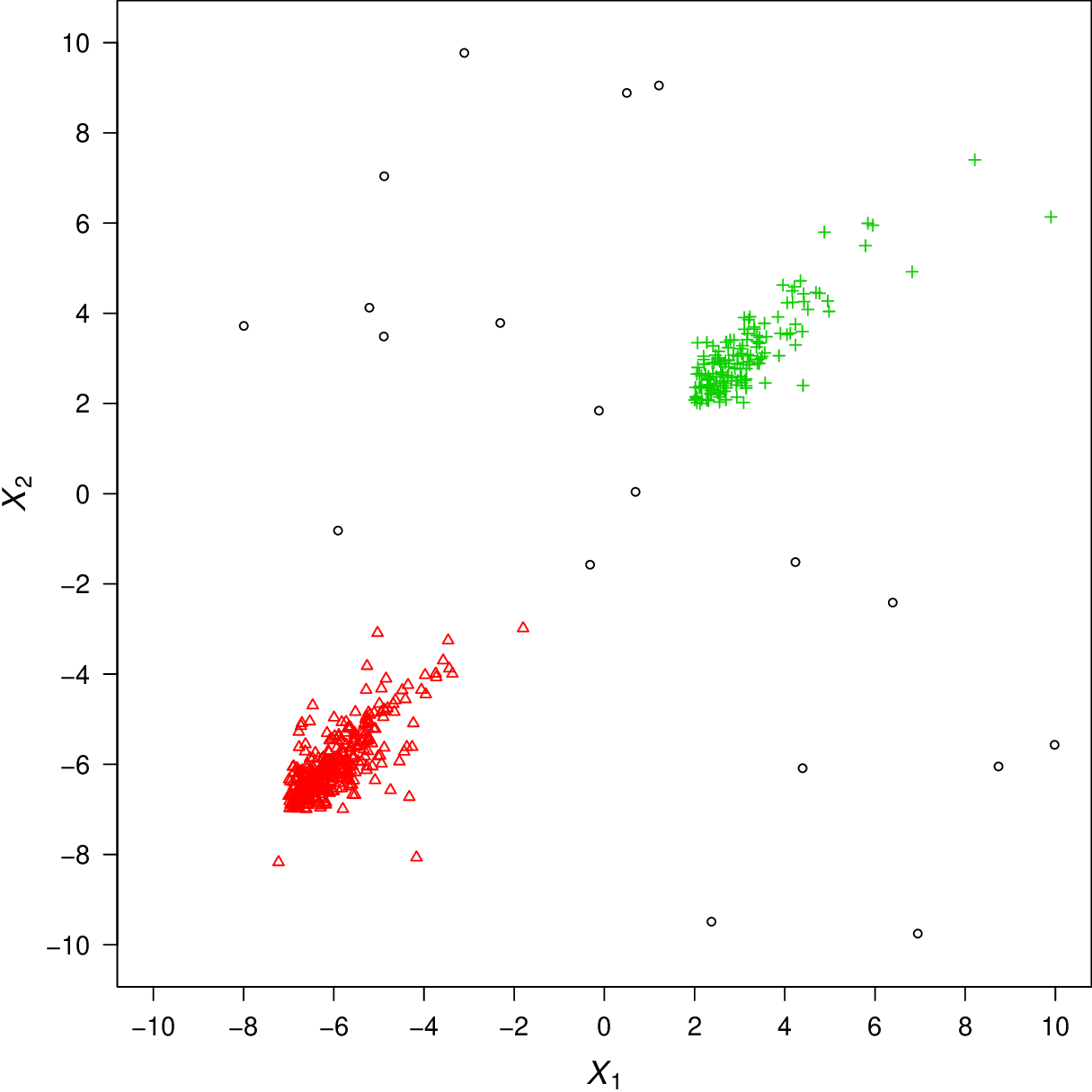}}}
%\quad
\subfigure[SCN mixture with $G=3$\label{fig:SCN.noisescatter1}]
{\resizebox{0.3\textwidth}{!}{\includegraphics{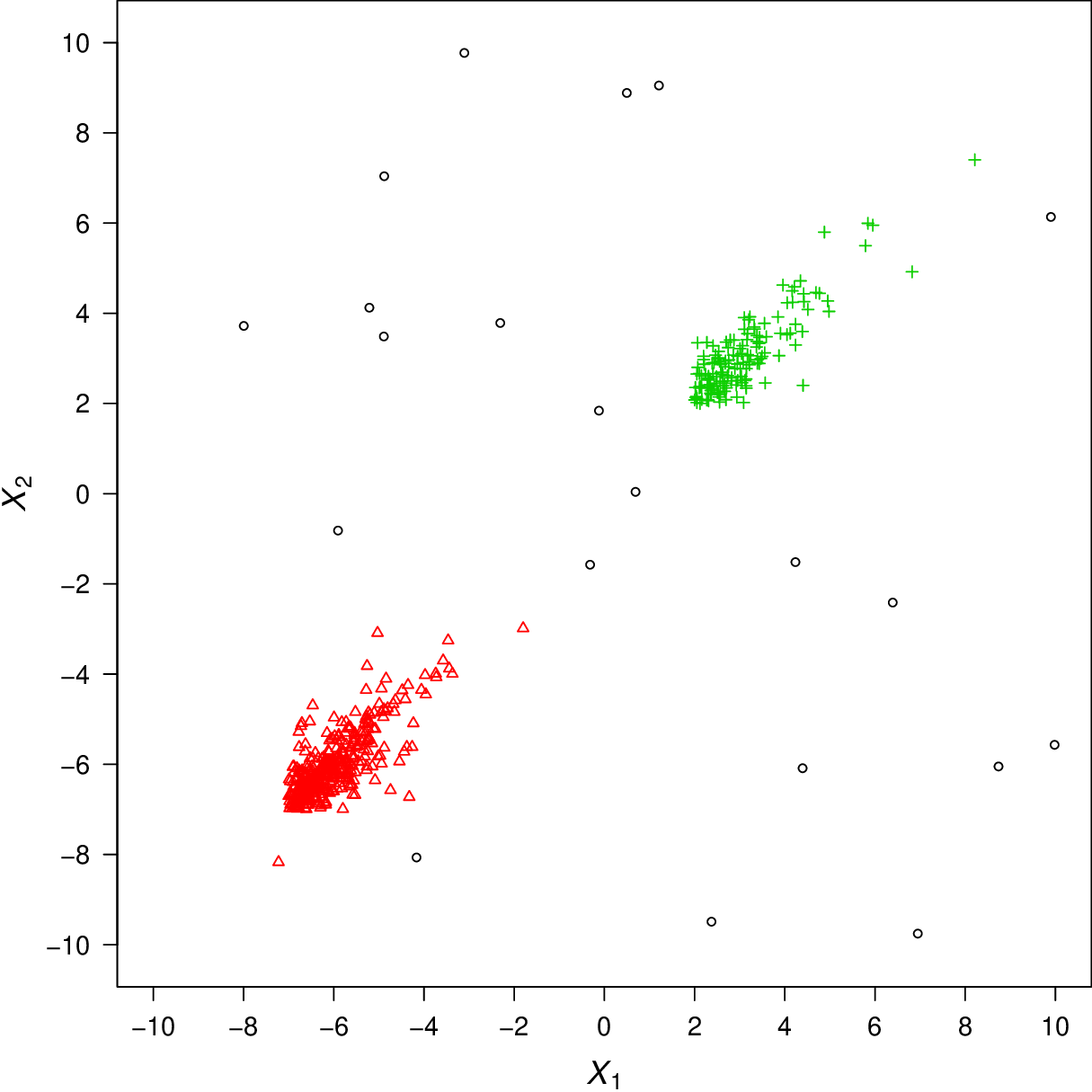}}}
%%%%%%%%%%%%
%% Others %%
%%%%%%%%%%%%
\subfigure[SS mixture with $G=3$\label{fig:SS.noisescatter1}]
{\resizebox{0.3\textwidth}{!}{\includegraphics{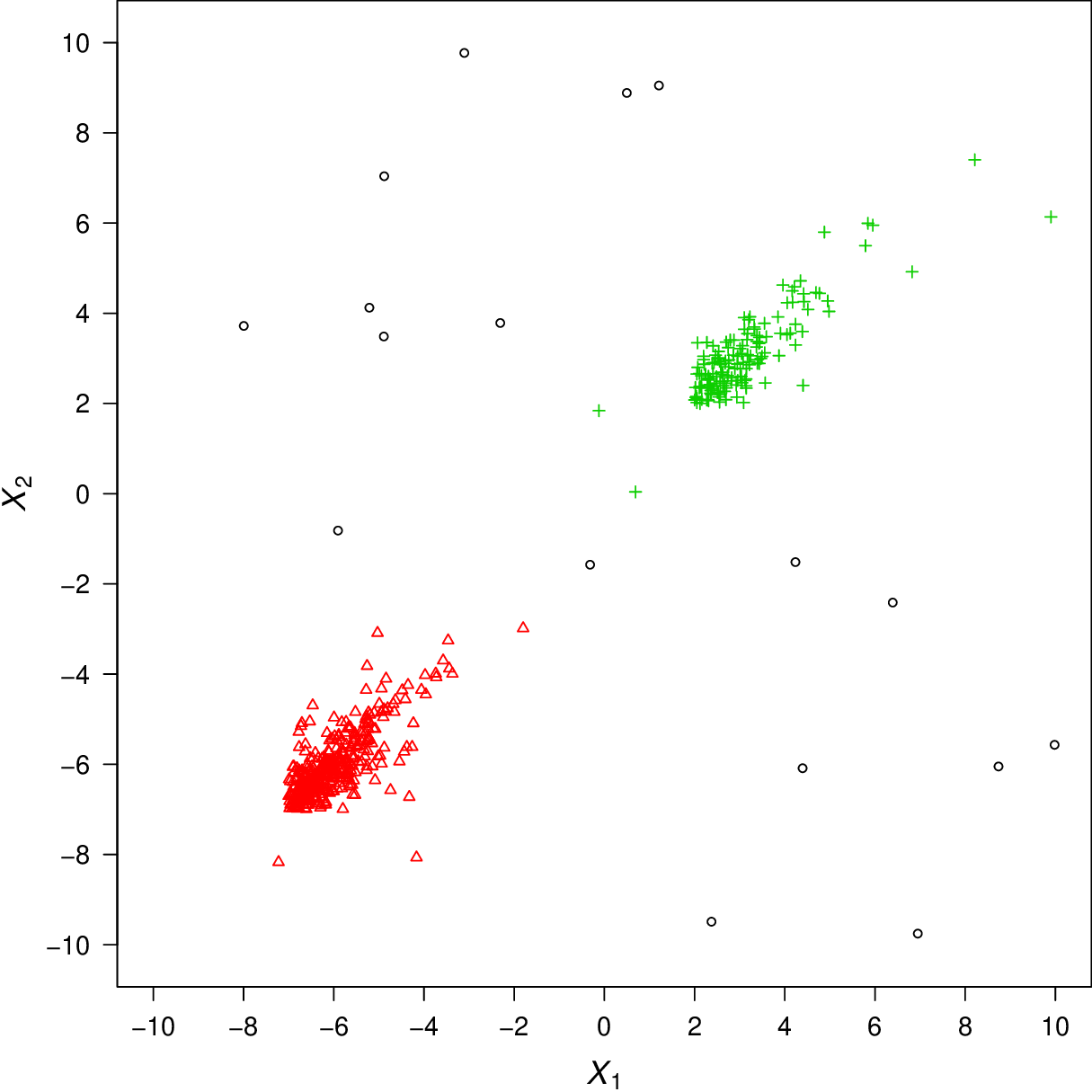}}}
%\quad
\subfigure[SAL mixture with $G=3$\label{fig:SAL.noisescatter1}]
{\resizebox{0.3\textwidth}{!}{\includegraphics{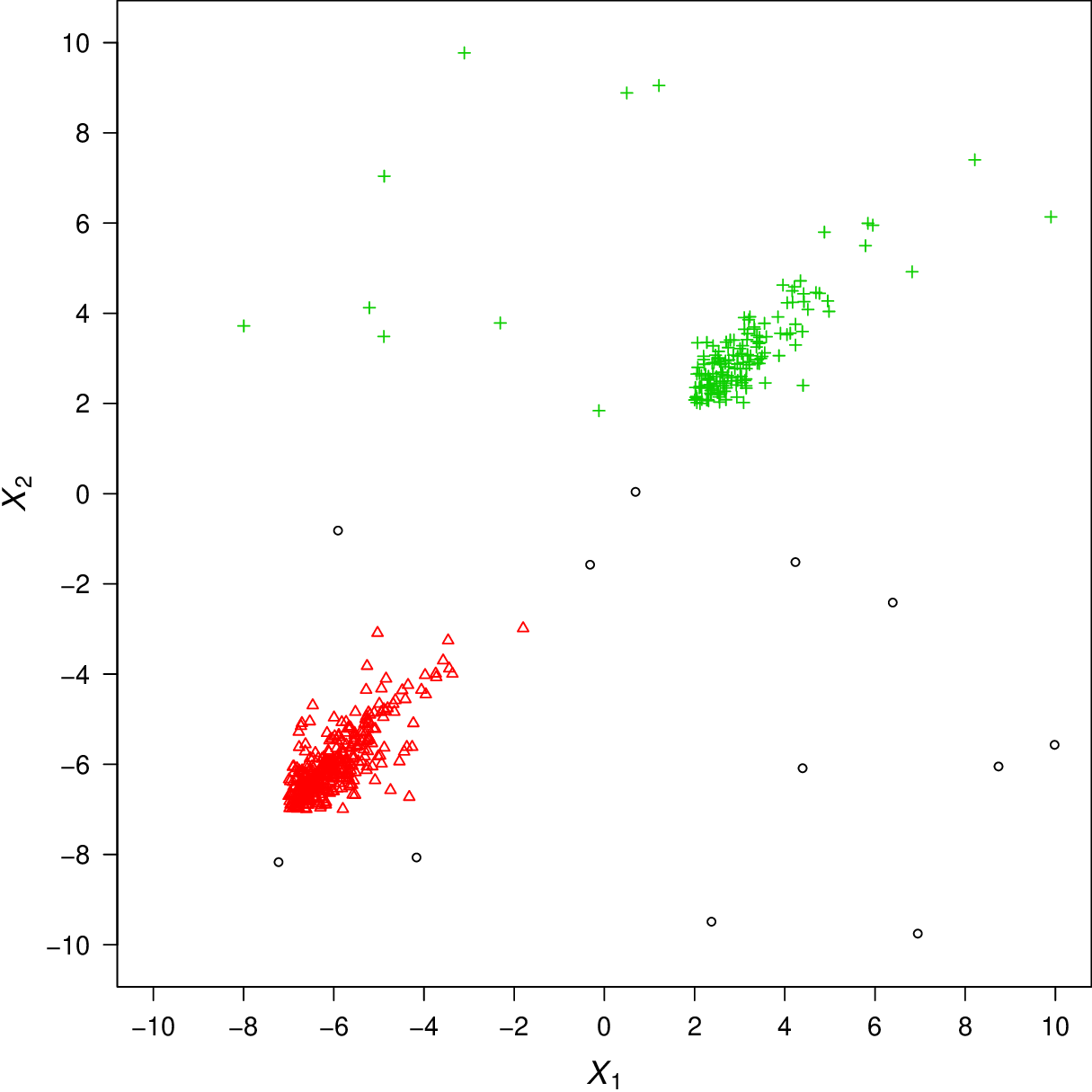}}}
%\quad
\subfigure[CSAL mixture with $G=2$\label{fig:CSAL.noisescatter1}]
{\resizebox{0.3\textwidth}{!}{\includegraphics{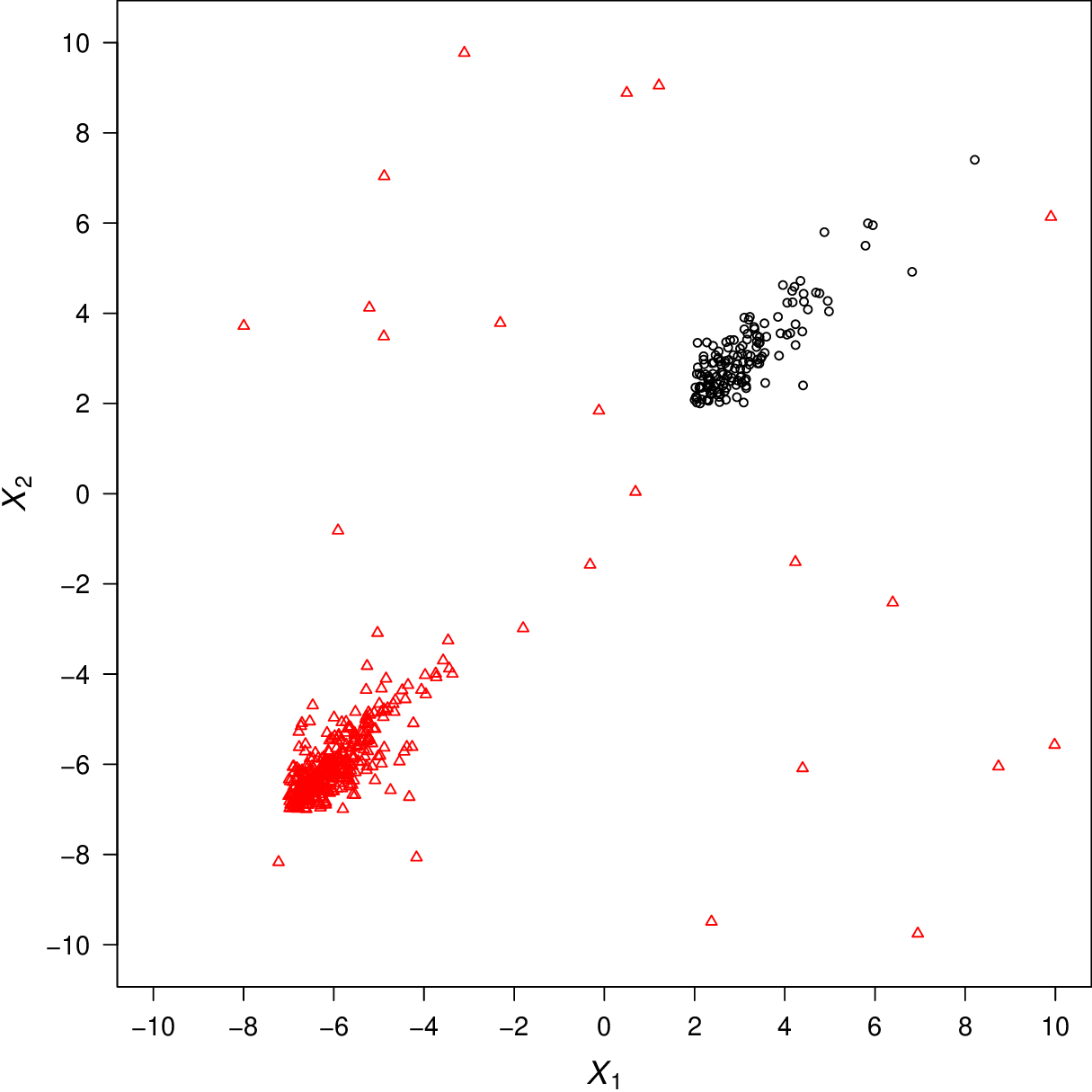}}}
\caption{
Simulated data from Section~\ref{subsubsec: two groups and two dimensions}. 
Scatter plots and MAP-classification of the observations from the models selected by the BIC.
\label{fig:noisescatter1 - best BIC}
}
\end{figure}

The competing models are fitted to the generated data for $G\in\left\{1,2,3,4\right\}$, resulting in a total of 36 fitted models.
The corresponding BIC values are reported in \tablename~\ref{tab:example 1 - BIC}.
The best (highest) BIC value for each family of models is highlighted in bold, while the overall best BIC value is highlighted in bold-italic. 
\begin{table}[!ht]
\caption{
Simulated data from Section~\ref{subsubsec: two groups and two dimensions}. 
BIC values for the fitted models.
The best BIC value for each column is highlighted in bold, while the overall best in bold-italic.  
}
\label{tab:example 1 - BIC}
\centering
\resizebox{\textwidth}{!}{
\begin{tabular}{c c rrrrrrrrrrr}
  \toprule
  %\backslashbox{$G$}{Mixture component} 	
	$G$ &&	N	&	$t$	&	CN	&	SN	&	S$t$	&	SCN	&	SS	&	SAL	&	CSAL	\\	
	\midrule
$1$	&	&	-5253.407	&	-4186.751	&	-4181.329	&	-4901.628	&	-3727.904	&	-3816.015	&	-3894.781	&	-4060.942	&	-3755.484	\\	
$2$	&	&	-3564.567	&	-2977.244	&	-3041.134	&	-3464.250	&	-2862.409	&	-2910.145	&	-3015.825	&	-3084.261	&	\textit{\textbf{-2829.773}}	\\	
$3$	&	&	-2969.665	&	-2931.081	&	-2990.333	&	-2918.205	&	\textbf{-2854.732}	&	\textbf{-2834.897}	&	\textbf{-2940.549}	&	\textbf{-2991.041}	&	-2886.742	\\	
$4$	&	&	\textbf{-2887.413}	&	\textbf{-2894.023}	&	\textbf{-2936.557}	&	\textbf{-2858.626}	&	-2861.371	&	-2890.005	&	-2971.456	&	-3005.068	&	-2927.472	\\	
   \bottomrule
\end{tabular}
}
\end{table}
For each considered family, the scatter plot of the best model according to the BIC, along with the MAP-classification of the observations, is graphically represented in \figurename~\ref{fig:noisescatter1 - best BIC}.
The BIC selects $G=4$ clusters for the mixtures with elliptically symmetric (N, $t$ and CN) components.
For these models, one component fits the cluster on the top-right, two components are needed to reproduce the asymmetric shape of the cluster on the bottom-left, and another component is attempting to model the background noise (see~\figurename s~\ref{fig:N.noisescatter1}--\ref{fig:CN.noisescatter1}).
Surprisingly, the BIC selects $G=4$ components for the SN mixture too (see~\figurename~\ref{fig:SN.noisescatter1}); the MAP-classification of the observations from this model is analogous to the previous ones meaning that the empirical skewness of the cluster on the bottom-left is not well-reproduced by the skewness of a single SN distribution.
Apart from the CSAL mixture, the BIC selects $G=3$ clusters for all the other (S$t$, SCN, SS and SAL) mixtures.
By looking at \figurename s~\ref{fig:St.noisescatter1}--\ref{fig:SAL.noisescatter1}, two components of these mixtures fit the two clusters, as well as part of the background noise (although with a different extent depending on the model), and the additional component models the remaining part of the background noise.
Finally, the BIC selects $G=2$ components for the CSAL mixture; for this model, the tails of the component on the bottom-left capture a great part of the background noise (see~\figurename~\ref{fig:CSAL.noisescatter1}).
The mixture of two CSAL distributions is, overall, the best fitted model according to the BIC (see \tablename~\ref{tab:example 1 - BIC}). 

In terms of classification performance, we first analyze the behavior of the competing models in classifying the good data and then we evaluate the ability of the mixtures with contaminated components in detecting the background noise. 
As concerns the first part of this analysis, \tablename~\ref{tab:confusion matrices ex1} reports the confusion matrices between the true and predicted MAP-classification from each model selected by the BIC.
Note that confusion matrices are computed only with respect to the 500 true good observations. 
\begin{table}[!ht]
\caption{
Simulated data from Section~\ref{subsubsec: two groups and two dimensions}.
Confusion matrices, with respect to the good data only, from the models selected by the BIC.
}
\label{tab:confusion matrices ex1}
\centering
\subtable[N mixture]
{
\label{tab:N classification ex1}
%\resizebox{!}{0.072\textheight}{
\begin{tabular}{l rrrr}
\toprule
       & \multicolumn{4}{c}{Fitted}   \\
True   & 1 & 2 & 3 & 4\\
\midrule
1 & 144 &   6 &   0 &   0 \\
2 &   0 &   3 & 103 & 244 \\
\bottomrule
\end{tabular}
%}
}
\quad
\subtable[$t$ mixture]
{
\label{tab:t classification ex1}
%\resizebox{!}{0.072\textheight}{
\begin{tabular}{l rrrr}
\toprule
       & \multicolumn{4}{c}{Fitted}   \\
True   &  1 &  2 &  3 &  4\\
\midrule
 1 & 144 &   6 &   0 &   0 \\
 2 &   0 &   3 & 106 & 241 \\
\bottomrule
\end{tabular}
%}
}
\quad
\subtable[CN mixture]
{
\label{tab:CN classification ex1}
%\resizebox{!}{0.072\textheight}{
\begin{tabular}{l rrrr}
\toprule
       & \multicolumn{4}{c}{Fitted}   \\
True   &  1 &  2 &  3 &  4\\
\midrule
 1 & 144 &   6 &   0 &   0 \\
 2 &   0 &   3 & 116 & 231 \\
\bottomrule
\end{tabular}
%}
}
\subtable[SN mixture]
{
\label{tab:SN classification ex1}
%\resizebox{!}{0.072\textheight}{
\begin{tabular}{l rrrr}
\toprule
       & \multicolumn{4}{c}{Fitted}   \\
True   & 1 & 2 & 3 & 4\\
\midrule
1 & 148 &   2 &   0 &   0 \\
2 &   0 &   3 & 121 & 226 \\
\bottomrule
\end{tabular}
%}
}
\quad
\subtable[S$t$ mixture]
{
\label{tab:St classification ex1}
%\resizebox{!}{0.072\textheight}{
\begin{tabular}{l rrr}
\toprule
       & \multicolumn{3}{c}{Fitted}   \\
True   &  1 &  2 &  3 \\
\midrule
 1 & 150 &   0 &   0 \\
 2 &   0 &   2 & 348 \\
\bottomrule
\end{tabular}
%}
}
\quad
\subtable[SCN mixture]
{
\label{tab:SCN classification ex1}
%\resizebox{!}{0.072\textheight}{
\begin{tabular}{l rrr}
\toprule
       & \multicolumn{3}{c}{Fitted}   \\
True   &  1 &  2 &  3 \\
\midrule
 1 & 150 &   0 &   0 \\
 2 &   0 &   2 & 348 \\
\bottomrule
\end{tabular}
%}
}\\
\subtable[SS mixture]
{
\label{tab:SS classification ex1}
%\resizebox{!}{0.072\textheight}{
\begin{tabular}{l rrr}
\toprule
       & \multicolumn{3}{c}{Fitted}   \\
True   &  1 &  2 &  3 \\
\midrule
 1 & 150 &   0 &   0 \\
 2 &   1 &   1 & 348 \\
\bottomrule
\end{tabular}
%}
}
\quad
\subtable[SAL mixture]
{
\label{tab:SAL classification ex1}
%\resizebox{!}{0.072\textheight}{
\begin{tabular}{l rrr}
\toprule
       & \multicolumn{3}{c}{Fitted}   \\
True   &  1 &  2 &  3 \\
\midrule
 1 & 150 &   0 &   0 \\
 2 &   0 &   2 & 348 \\
\bottomrule
\end{tabular}
%}
}
\quad
\subtable[CSAL mixture]
{
\label{tab:CSAL classification ex1}
%\resizebox{!}{0.072\textheight}{
\begin{tabular}{l rr}
\toprule
       & \multicolumn{2}{c}{Fitted}   \\
True   &  1 &  2 \\
\midrule
 1 & 150 &   0 \\
 2 &   0 & 350 \\
\bottomrule
\end{tabular}
%}
}
\end{table}
We highlight that the two underlying groups of good data are classified correctly 
%(in terms of number of mixture components and misclassifications) 
by the CSAL mixture only ($\text{ARI}=1$); see \tablename~\ref{tab:CSAL classification ex1}.
The selected N mixture roughly splits the second group in two sub-clusters, while the cluster devoted to the background noise also absorbs 6 observations from group 1, and 3 observations from group 2 ($\text{ARI}=0.584$); see \tablename~\ref{tab:N classification ex1}.
Similar results are obtained for $t$, CN and SN mixtures, with ARI values equal to 0.578, 0.560 and 0.561, respectively; see \tablename s~\ref{tab:t classification ex1}--\ref{tab:SN classification ex1}.
For the mixture models with $G=3$ skewed (S$t$, SCN, SS and SAL) clusters, selected by the BIC, the first group is classified correctly, in the sense that it is not split by the fitted models in sub-groups; however, some (one or two) of the observations in the second group are assigned to the noise component of these models and, for the SS mixture, one observation from group 2 is joined to the first component of the mixture.
The ARI values for these models are high (0.989 for S$t$, SCN and SAL mixtures, and 0.986 for the SS mixture).
The CSAL mixture selected by the BIC is the best performer, in terms of classification of the good data, being the only one producing a correct number of mixture components and a perfect classification ($\text{ARI}=1$).  

For the purpose of evaluation of the performance of the competing mixtures of contaminated distributions in detecting the background noise, we report the true positive rate (TPR), measuring the proportion of bad points that are correctly identified as bad points, and the false positive rate (FPR), corresponding to the proportion of good points incorrectly classified as bad points.
We consider the fitted CN, SCN, and CSAL mixtures with $G=2$ components. 
We use the outlier detection rule detailed in \citet{Punz:McNi:Robu:2016} for the CN mixture, and the detection rule illustrated in Section~\ref{subsec:Automatic detection of bad points} for the CSAL mixture.
The underlying principle of these rules is the same.
We apply an analogous outlier detection rule for the SCN mixture; to our knowledge, this is the first time such a detection rule is used on the SCN mixture proposed by \citet{cabral12}.   
\tablename~\ref{tab:TPR and FPR ex1} reports the obtained TPRs and FPRs from the three model-based outlier detection rules.
\begin{table}[!ht]
\caption{Simulated data from Section~\ref{subsubsec: two groups and two dimensions}. 
TPRs and FPRs from the CN, SCN and CSAL mixtures with $G=2$ components.}
\centering{
\begin{tabular}{l ccc}
\toprule
 	  &	CN mixture	&	SCN mixture	&	CSAL mixture	\\	
		\midrule
TPR	&	0.920	&	0.880	&	0.800	\\	
FPR	&	0.106	&	0.032	&	0.004	\\	
\bottomrule
\end{tabular}
}
\label{tab:TPR and FPR ex1}
\end{table}

\begin{figure}[!ht]
\centering
\subfigure[CN mixture\label{fig:CN.noisedetect1}]
{\resizebox{0.3\textwidth}{!}{\includegraphics{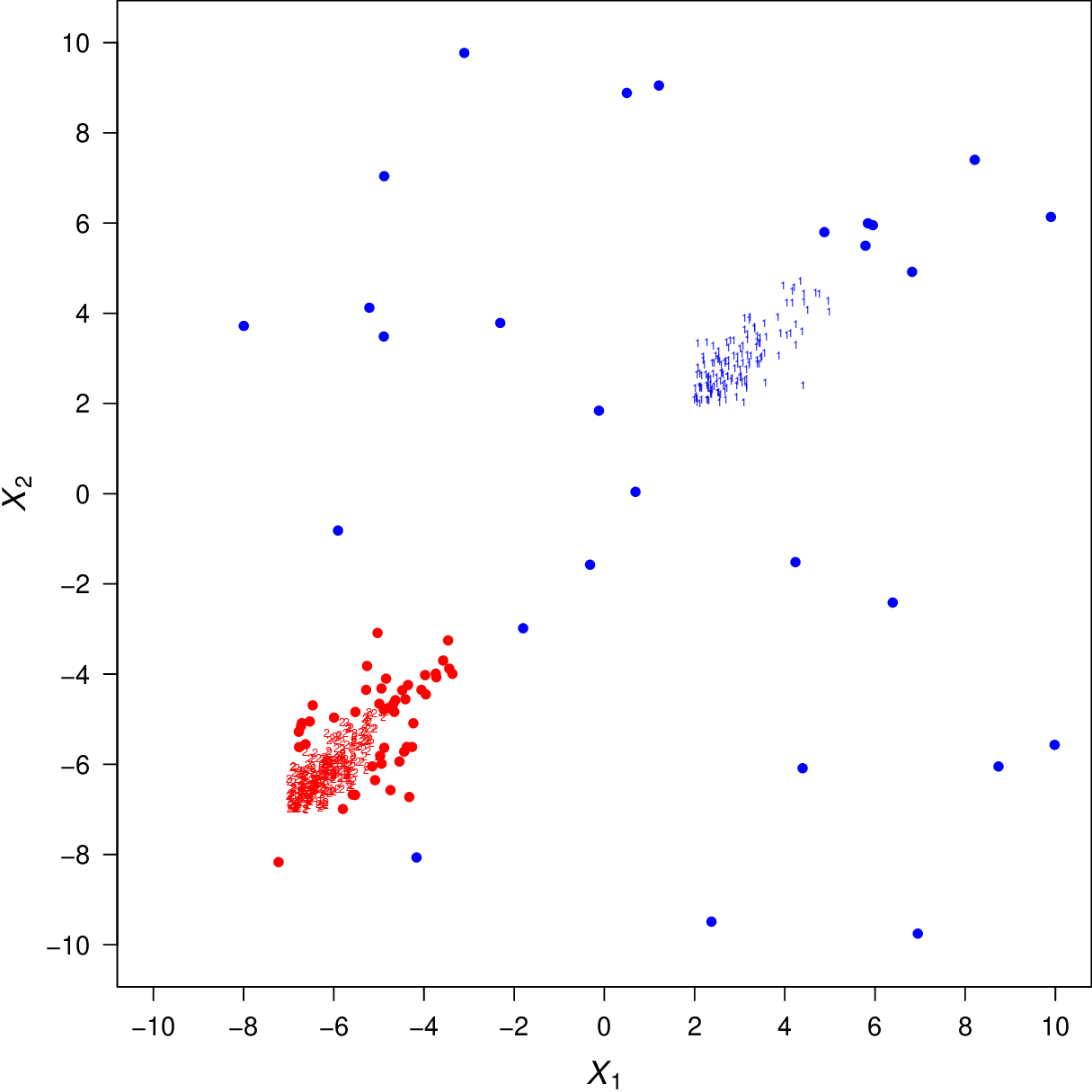}}}
%\quad
\subfigure[SCN mixture\label{fig:SCN.noisedetect1}]
{\resizebox{0.3\textwidth}{!}{\includegraphics{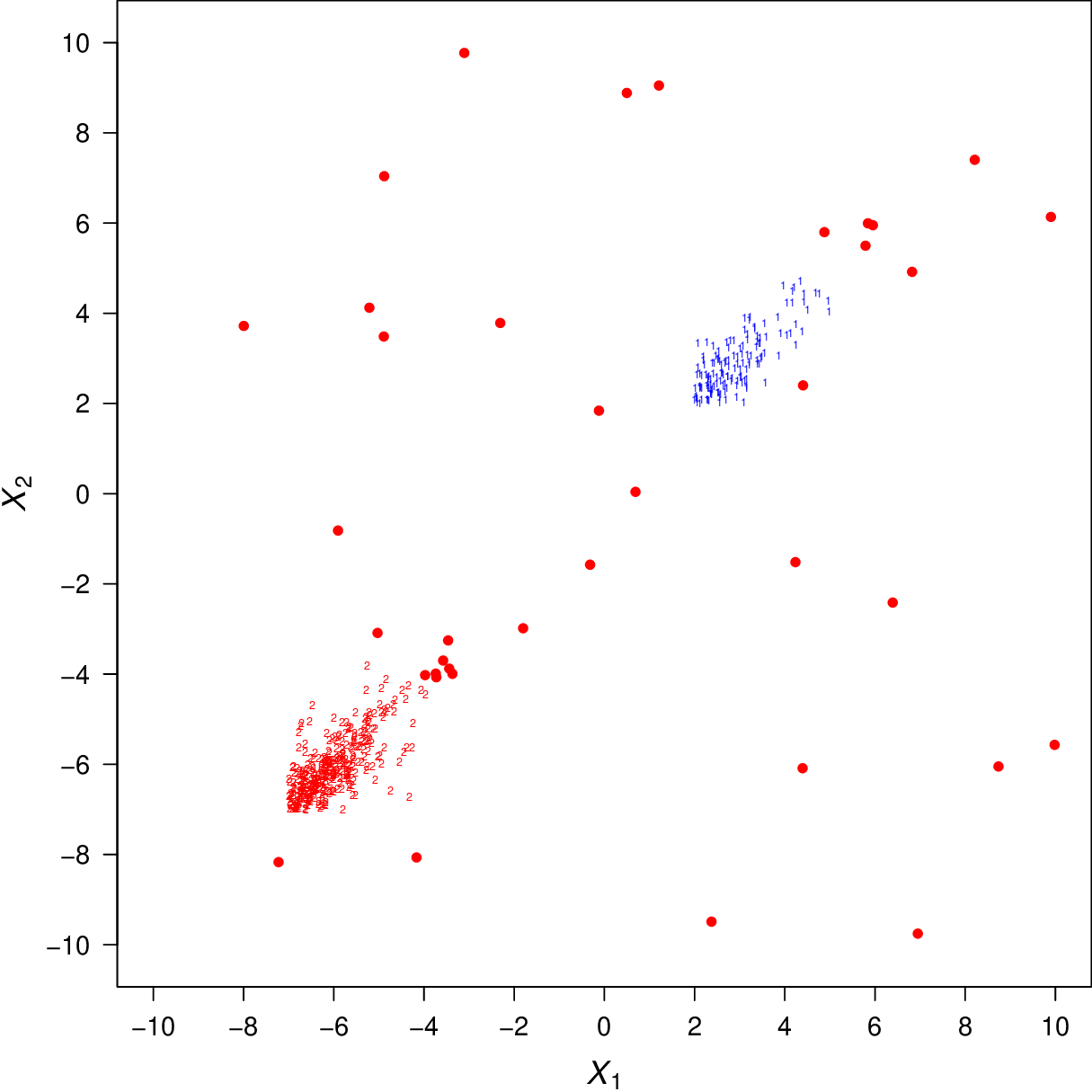}}}
%\quad
\subfigure[CSAL mixture\label{fig:CSAL.noisedetect1}]
{\resizebox{0.3\textwidth}{!}{\includegraphics{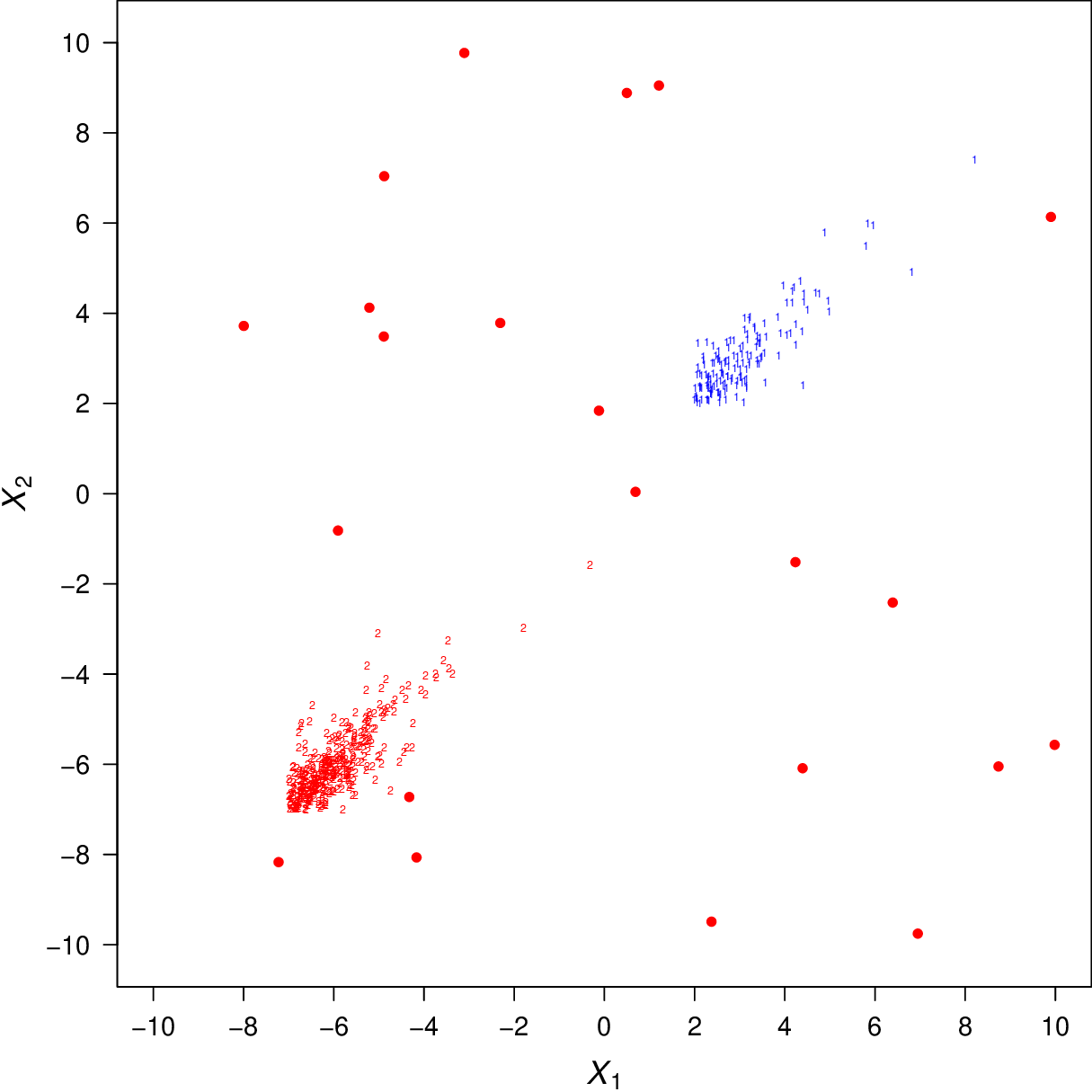}}}
\caption{
Simulated data from Section~\ref{subsubsec: two groups and two dimensions}. 
Scatter plots and MAP-classification of the observations from the CN, SCN and CSAL mixtures with $G=2$ clusters.
Bullets denote detected bad points.
\label{fig:Outlier detection ex1}
}
\end{figure}

The CN mixture shows the highest TPR, while the CSAL mixture gives the lowest (almost optimal) FPR.
To find out more about the motivation of these results, \figurename~\ref{fig:Outlier detection ex1} shows, for each of the three mixtures of contaminated distributions, the scatter plot of the observations with symbols and colors diversified with respect to the group MAP-membership; detected outliers are denoted by bullets.
The FPR from the CN mixture is high (0.106) because a lot of good observations on the boundary of the bottom-left group are seen as outliers by the CN density of that cluster.
In detail, an ``elliptical'' subset of the group is accommodated by the good component of the CN density, while the rest is banished to the bad one.
% as formed by good data while the remaining part of the component is seen as bad.
For this CN component, the estimated proportion of good points is 0.778 while the corresponding degree of contamination is 5.824.
By comparing \figurename~\ref{fig:noisescatter1} with \figurename~\ref{fig:CN.noisedetect1}, the CN mixture behaves in a similar way, even if with a lower extent, on the top-right group.
However, the bad component of the CN distribution is now more devoted to capture the background noise.
This is confirmed by an high estimated degree of contamination (20.889), which is greater than the degree of contamination (5.824) of the other CN mixture component; the estimated proportion of good observations is now 0.811.
The FPR, equal to 0.032, from the SCN mixture is due to the fact the long north-east tail of both the groups of good data is instead seen as composed by outliers by the model. 
The ``low'' TPR from the CSAL mixture is 0.800 is mainly due to two reasons.
Firstly, some of the added noise points end up in the interior of the groups (see~\figurename~\ref{fig:noisescatter1}), making it difficult to identify them as outliers by any detection rule.
Secondly, the two good SAL components of the two CSAL distributions of the mixture are, by theirself, already able to cope with a certain fraction of the background noise.
The remaining part of the noise is captured by the bad SAL component of the CSAL distribution located on the bottom-left. 
This distribution has an estimated proportion of good data equal to $\hat{\lambda}_2=0.940$ and a high estimated degree of contamination ($\hat{\rho}_2=331.409$).

\subsubsection{Example with $G=3$ groups and $p=2$ dimensions}
\label{subsubsec: three groups and two dimensions}

\begin{figure}[!ht]
\centering
\resizebox{0.4\textwidth}{!}{
\includegraphics{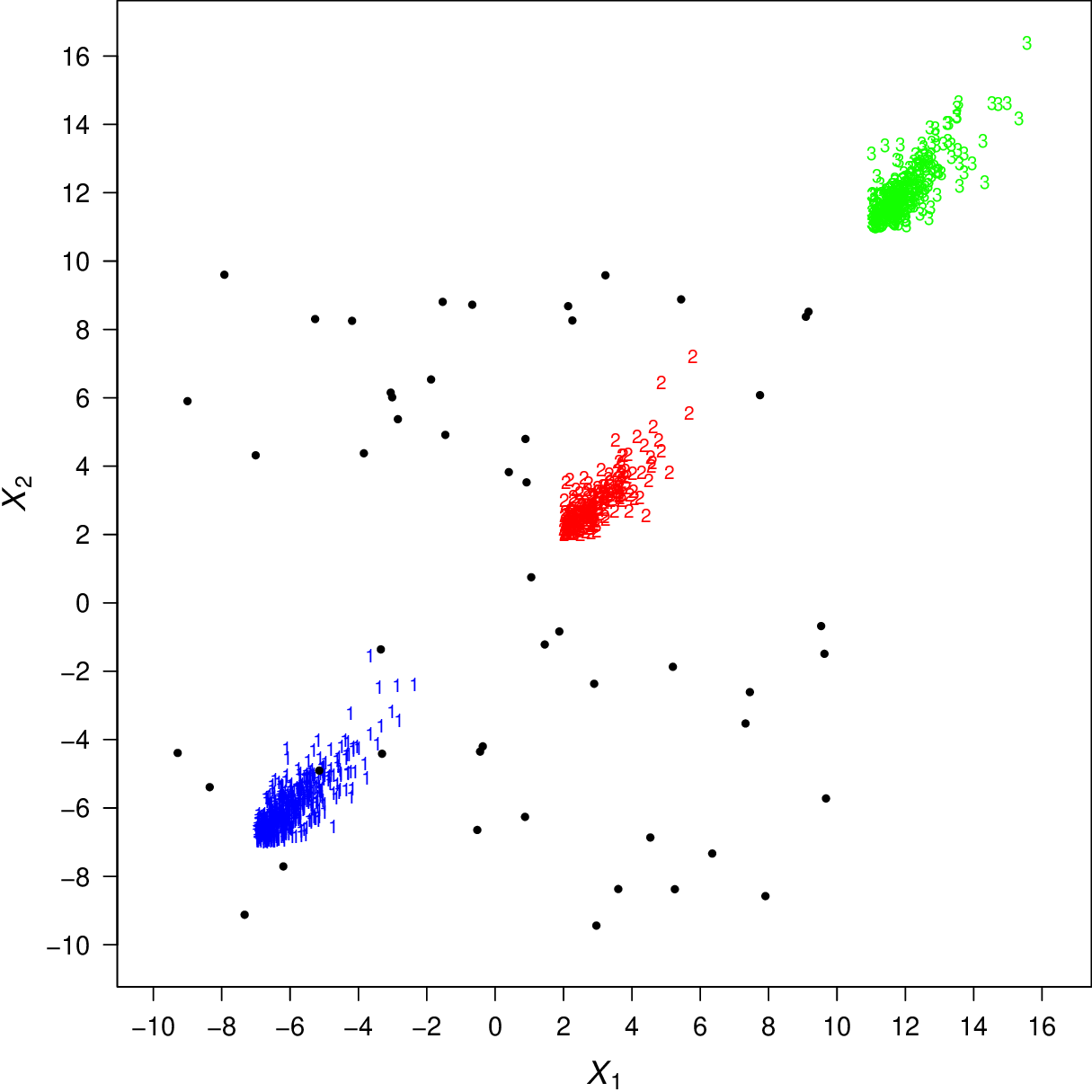} 
}
\caption{
Simulated data from Section~\ref{subsubsec: three groups and two dimensions}. 
Scatter plot (\textcolor{black}{1}, \textcolor{red}{2} and \textcolor{green}{3} denote the first, second and third group, respectively).
Background uniform noise points are denoted by $\bullet$.
}
\label{fig:noisescatter3}
\end{figure}

In the second example, $n=1000$ bivariate ($p=2$) observations are generated by a mixture of $G=3$ $t$-slice distributions with parameters
\begin{equation*}
	\pi_1=0.5,
	\ 
	\pi_2=0.2,
	\ 
	\bmu_1=
	\begin{bmatrix*}[c]
-7	\\
-7	 
\end{bmatrix*},
\ 
	\bmu_2=
	\begin{bmatrix*}[c]
2	\\
2	 
\end{bmatrix*},
\ 
	\bmu_3=
	\begin{bmatrix*}[c]
11	\\
11	 
\end{bmatrix*},
\ \bSigma_1=\bSigma_2=\bSigma_3=\begin{bmatrix*}[c]
1 & 0.9	\\
0.9 & 1	 
\end{bmatrix*},
\ \text{and} \ 
	\nu_1=\nu_2=\nu_3=6.
%\label{eq:common generating parameters}
\end{equation*}
As for the the example in Section~\ref{subsubsec: two groups and two dimensions}, 50 noise points are added from a uniform distribution over the interval $(-10, 10)$ on each dimension.
%This yields an overall dataset comprising $n=525$ observations.
The scatter plot of the generated data, with labels \textcolor{black}{1}, \textcolor{red}{2} and \textcolor{green}{3} for points from the first, second and third group, respectively, and with bullets denoting uniform noise points, is displayed in \figurename~\ref{fig:noisescatter3}. 
As we can see, in this example the noise covers only two of the three generated clusters. 

The competing models are fitted to the generated data for $G\in\left\{1,2,3,4,5\right\}$.
The corresponding BIC values are reported in \tablename~\ref{tab:example 3 - BIC}.
%The best (highest) BIC value for each family of models is highlighted in bold, while the overall best BIC value is highlighted in bold-italic. 
\begin{table}[!ht]
\caption{
Simulated data from Section~\ref{subsubsec: three groups and two dimensions}. 
BIC values for the fitted models.
The best BIC value for each column is highlighted in bold, while the overall best in bold-italic.  
}
\label{tab:example 3 - BIC}
\centering
\resizebox{\textwidth}{!}{
\begin{tabular}{c c rrrrrrrrrrr}
  \toprule
  %\backslashbox{$G$}{Mixture component} 	
	$G$ &&	N	&	$t$	&	CN	&	SN	&	S$t$	&	SCN	&	SS	&	SAL	&	CSAL	\\	
	\midrule
$1$	&	&	-11698.509	&	-9947.977	&	-9659.884	&	-11138.797	&	-9172.015	&	-9076.360	&	-9675.003	&	-10634.259	&	-10036.071	\\
$2$	&	&	-9180.156	&	-8143.511	&	-7889.729	&	-8838.385	&	-7687.617	&	-7944.347	&	-8220.684	&	-8167.903	&	-7629.978	\\
$3$	&	&	-7607.284	&	-6788.780	&	-6850.640	&	-7340.803	&	\textbf{-6534.969}	&	\textbf{-6530.765}	&	-6647.791	&	-6777.626	&	\textbf{\textit{-6435.498}}	\\
$4$	&	&	-7490.312	&	\textbf{-6634.094}	&	\textbf{-6654.045}	&	\textbf{-6487.831}	&	-6575.936	&	-6559.681	&	\textbf{-6473.617}	&	\textbf{-6459.348}	&	-6512.047	\\
$5$	&	&	\textbf{-6752.651}	&	-6651.128	&	-6737.504	&	-7385.226	&	-6608.633	&	-6549.832	&	-6504.393	&	-6468.838	&	-6535.544	\\
   \bottomrule
\end{tabular}
}
\end{table}

We can note how the best model, among the 45 models being fitted, is the CSAL mixture with $G=3$ clusters.
For the models selected by the BIC, \tablename~\ref{tab:example 3 - ERROR and ARI} reports the corresponding ARI values with respect to the good points only. 
\begin{table}[!ht]
\caption{
Simulated data from Section~\ref{subsubsec: three groups and two dimensions}. 
ARI values, on the good data only, for the mixture models selected by the BIC.
%Bold is used for the best value of $G$ for each model, while bold-italic is used for the overall best model.  
}
\label{tab:example 3 - ERROR and ARI}
\centering
%\resizebox{\textwidth}{!}{
\begin{tabular}{c c ccccccccccc}
  \toprule
  %\backslashbox{$G$}{Mixture component} 	
	 &&	N	&	$t$	&	CN	&	SN	&	S$t$	&	SCN	&	SS	&	SAL	&	CSAL	\\	
	\midrule
$G$	&	&	5	&	4	&	4	&	4	&	3	&	3	&	4	&	4	&	3	\\
%Error rate	&	&	0.018	&	0.003	&	0.004	&	0.013	&	0.000	&	0.004	&	0.002	&	0.403	&	0.000	\\
ARI	&	&	0.970	&	0.996	&	0.994	&	0.980	&	1.000	&	0.988	&	0.997	&	0.310	&	1.000	\\
   \bottomrule
\end{tabular}
%}
\end{table}
The S$t$ and CSAL mixtures attain a perfect classification ($\text{ARI}=1$), while the SAL mixture is the model with the worst classification results ($\text{ARI}=0.310$).   

\begin{figure}[!ht]
\centering
\subfigure[CN mixture\label{fig:CN.noisedetect3}]
{\resizebox{0.3\textwidth}{!}{\includegraphics{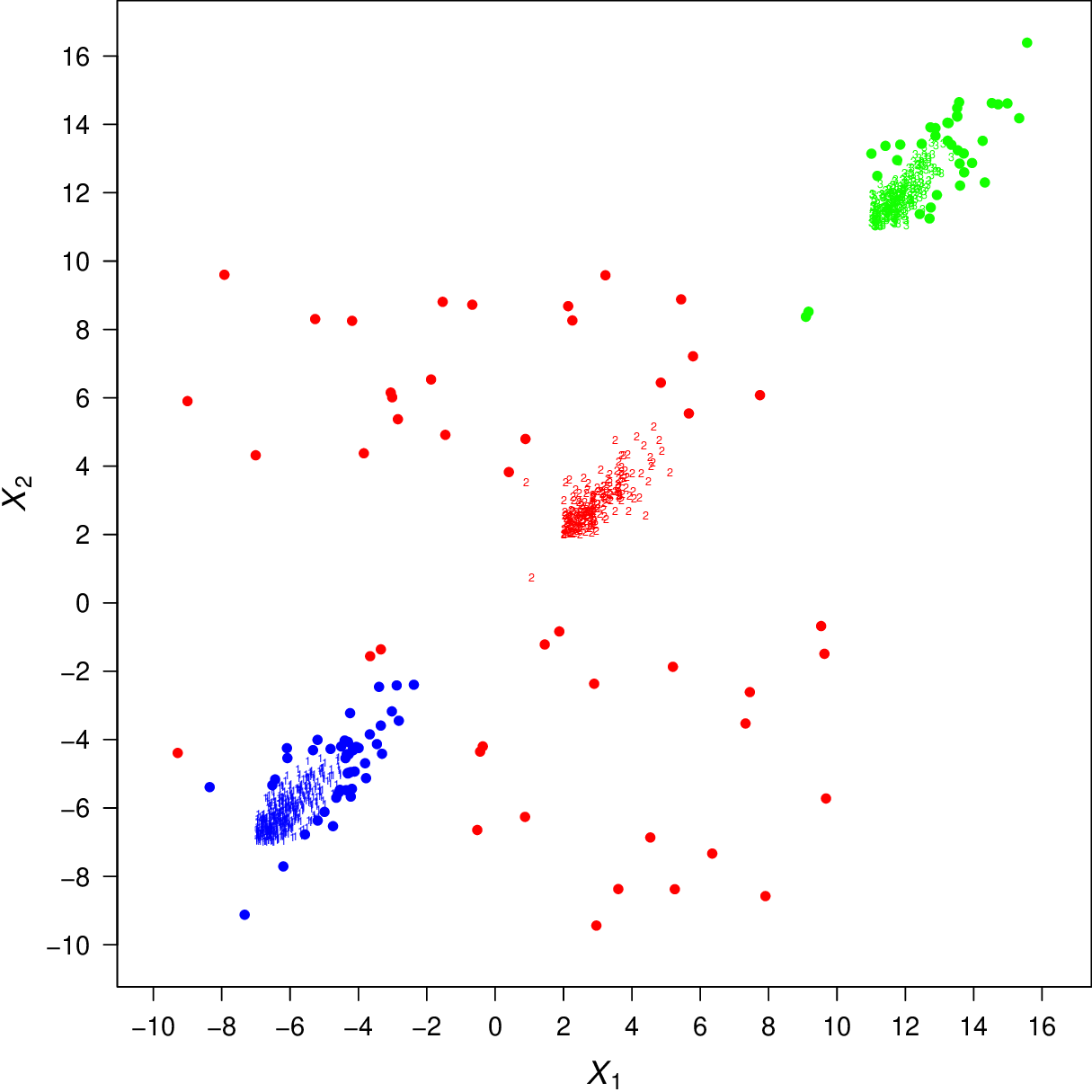}}}
%\quad
\subfigure[SCN mixture\label{fig:SCN.noisedetect3}]
{\resizebox{0.3\textwidth}{!}{\includegraphics{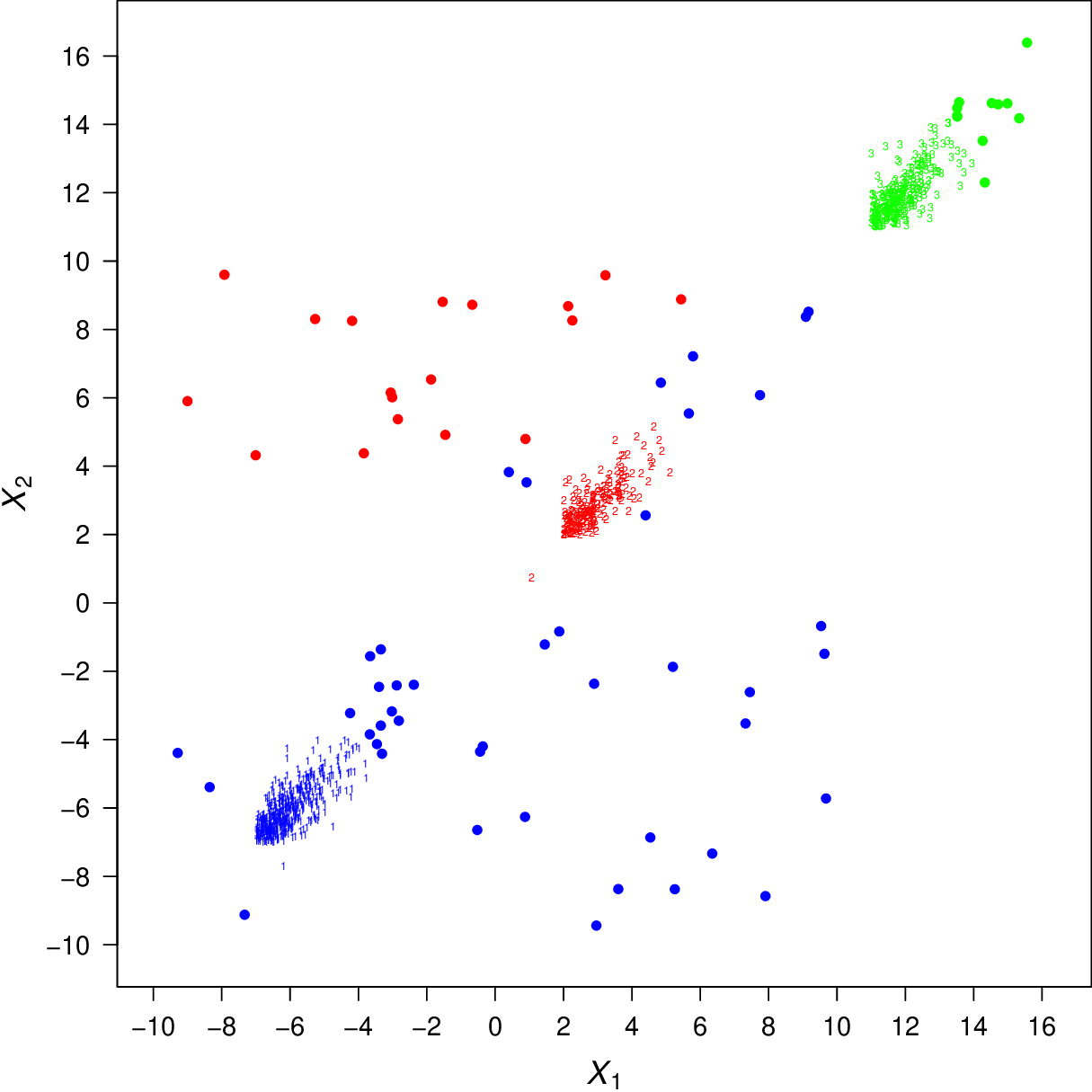}}}
%\quad
\subfigure[CSAL mixture\label{fig:CSAL.noisedetect3}]
{\resizebox{0.3\textwidth}{!}{\includegraphics{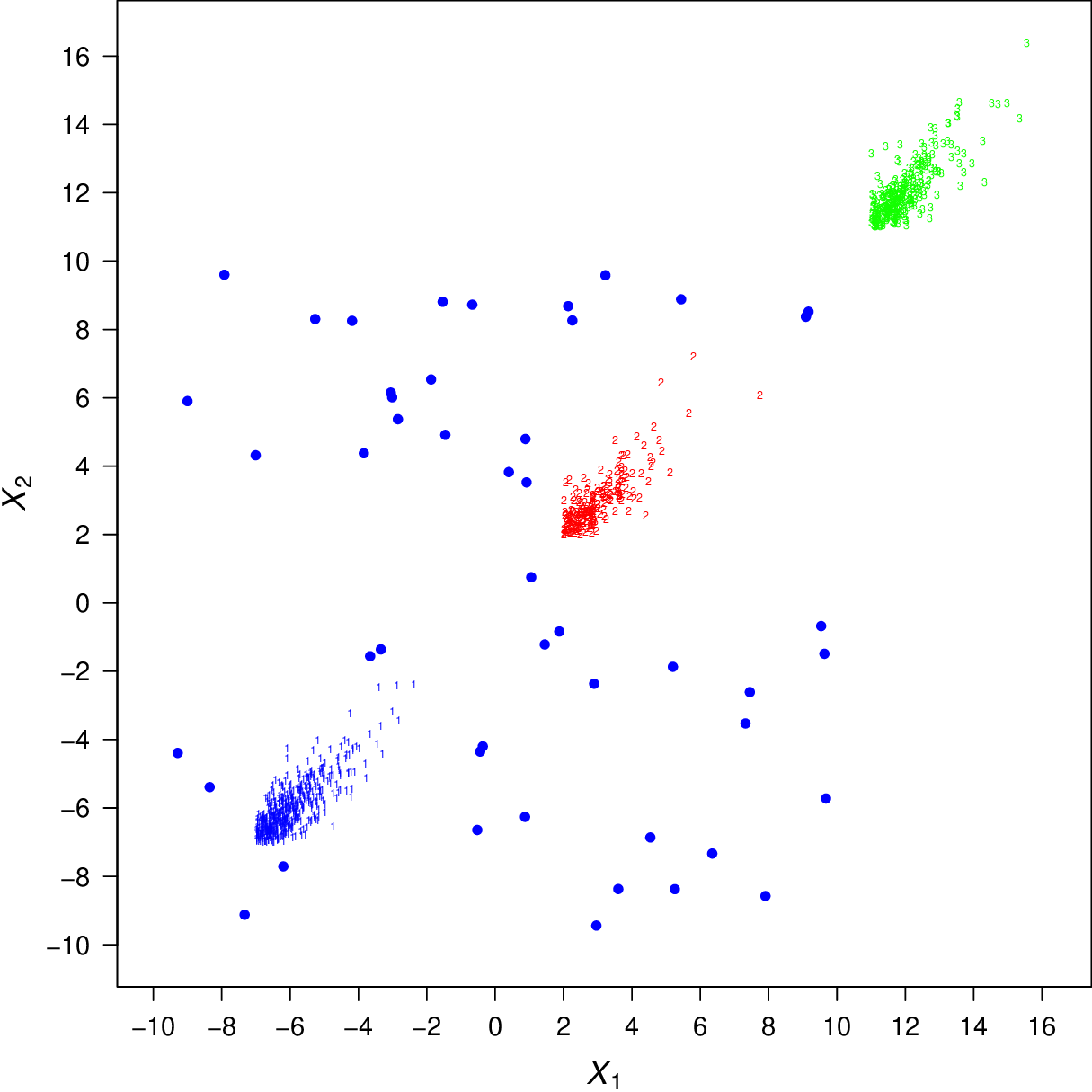}}}
\caption{
Simulated data from Section~\ref{subsubsec: three groups and two dimensions}. 
Scatter plots and MAP-classification of the observations from the CN, SCN and CSAL mixtures with $G=3$ clusters.
Bullets denote detected bad points.
\label{fig:Outlier detection ex3}
}
\end{figure}

\tablename~\ref{tab:TPR and FPR ex3} reports TPRs and FPRs from the application of the outlier detection rules from the fitted CN, SCN, and CSAL mixtures with $G=3$ clusters.
\begin{table}[!ht]
\caption{Simulated data from Section~\ref{subsubsec: three groups and two dimensions}. 
TPRs and FPRs from the CN, SCN and CSAL mixtures with $G=3$ components.}
%\vspace{-0.1cm}
\centering{
\begin{tabular}{l ccc}
\toprule
 	  &	CN mixture	&	SCN mixture	&	CSAL mixture	\\	
		\midrule
TPR	&	0.940	&	0.940	&	0.940	\\
FPR	&	0.080	&	0.025	&	0.001	\\
\bottomrule
\end{tabular}
}
\label{tab:TPR and FPR ex3}
\end{table}
The competing rules provide the same TPR (0.940), but the rule associated to the CSAL mixture is the one providing the best, almost perfect, FPR (0.001), followed by the rule associated to the SCN mixture ($\text{FPR}=0.025$). 
\figurename~\ref{fig:Outlier detection ex3} shows, for each of the three mixtures of contaminated distributions, the scatter plot of the observations with symbols and colors diversified with respect to the group MAP-membership; detected outliers are denoted by bullets.
As we can note, the considered models place their three components on the three underlying groups of good data.
However, their way to declare observations as outliers is different.
For the CN mixture, the bad component of the CN density on the middle accommodates a great part of the background noise.
Instead, the bad components of the two remaining CN densities accommodate the external part of the groups composed by good observations, and this contributes to increase the FPR for this model.
As concerns the SCN mixture, a part of the background noise is accommodated by the bad component of the SCN density on the bottom-left, while the remaining part is accommodated by the SCN density on the middle.
However, some good observations in the north-east direction of the groups on the bottom-left and the top-right are seen as outliers too, and this is the motivation of the non-optimal FPR for the detection rule from this model.
Instead, an almost perfect detecting behavior is observed for the CSAL mixture, with the bad component of the CSAL density on the bottom-left capturing all the background noise; with reference to this cluster of the CSAL mixture, the estimated mixture weight is $\hat{\pi}_1=0.524$, the estimated proportion of good data is $\hat{\lambda}_1=0.902$, and the estimated degree of contamination is $\hat{\rho}_1=182.088$.  

\subsubsection{Example with $G=2$ groups and $p=3$ dimensions}
\label{subsubsec: two groups and three dimensions}

\begin{figure}[!ht]
\centering
\resizebox{0.5\textwidth}{!}{
\includegraphics{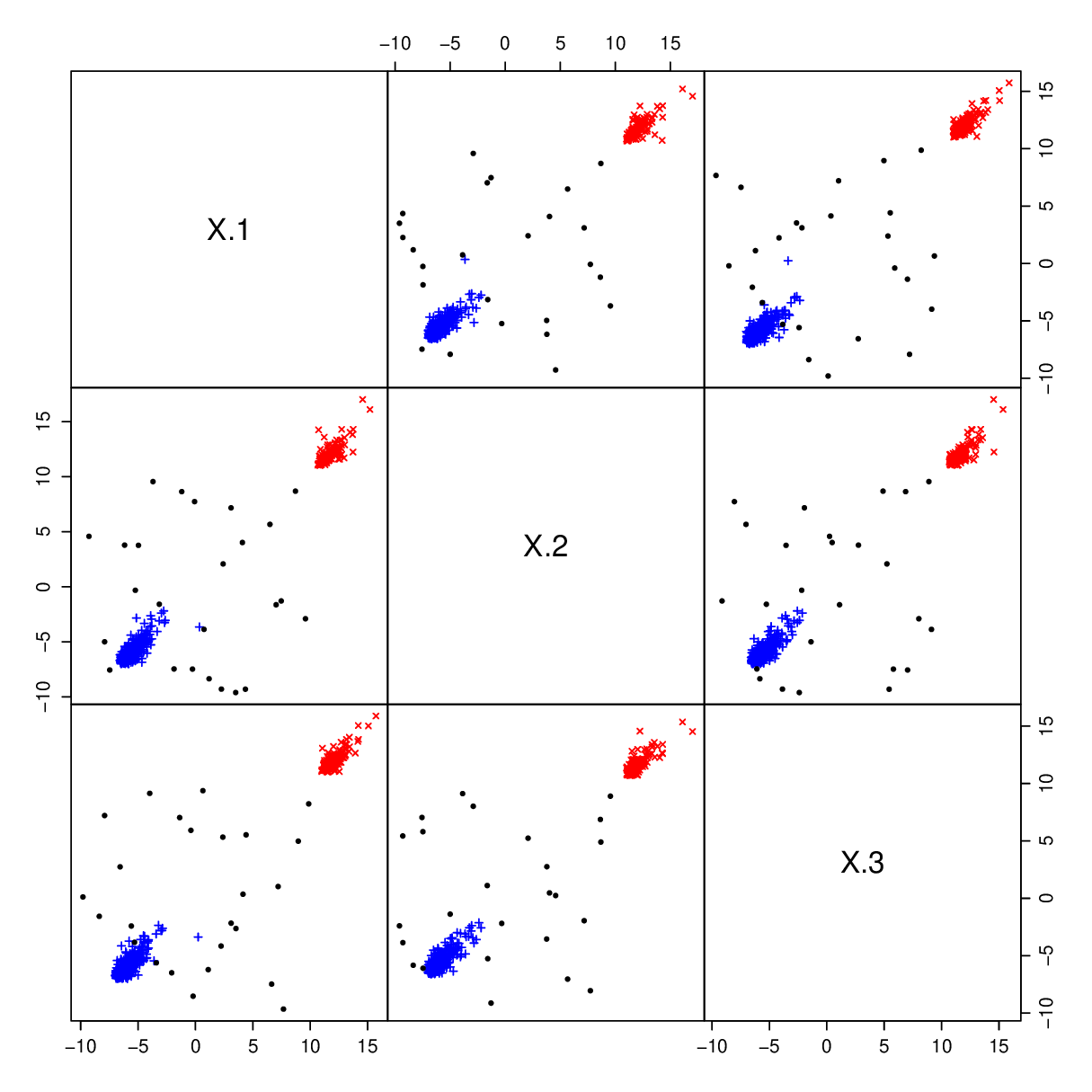} 
}
\caption{
Simulated data from Section~\ref{subsubsec: two groups and three dimensions}. 
Matrix of pairwise scatter plots (\textcolor{black}{$+$} and \textcolor{red}{$\times$} denote the first and second groups, respectively).
Background uniform noise points are denoted by $\bullet$.
}
\label{fig:noisescatter2}
\end{figure}

In the third example, $n=500$ trivariate ($p=3$) observations are generated by a mixture of $G=2$ $t$-slice distributions with parameters
\begin{equation*}
	\pi_1=0.7,
	\quad
	\bmu_1=
	\begin{bmatrix*}[c]
-7	\\
-7	\\
-7	 
\end{bmatrix*},
\quad 
	\bmu_2=
	\begin{bmatrix*}[c]
11	\\
11	\\
11	 
\end{bmatrix*},
\quad \bSigma_1=\bSigma_2=\begin{bmatrix*}[c]
1   & 0.9	& 0.9\\
0.9 & 1	  & 0.9\\
0.9 & 0.9	& 1 
\end{bmatrix*},
\quad \text{and} \quad 
	\nu_1=\nu_2=6.
%\label{eq:common generating parameters}
\end{equation*}
Moreover, 25 noise points are added from a uniform distribution over the interval $(-10, 10)$ on each dimension.
%This yields an overall dataset comprising $n=525$ observations.
The matrix of pairwise scatter plots of the generated data, with labels \textcolor{black}{1} and \textcolor{red}{2} for points from the first and second group, respectively, and with bullets denoting uniform noise points, is displayed in \figurename~\ref{fig:noisescatter2}. 
As we can see, the noise covers only one of the two generated clusters. 

The competing models are fitted to the generated data for $G\in\left\{1,2,3,4\right\}$.
The corresponding BIC values are reported in \tablename~\ref{tab:example 2 - BIC}.
\begin{table}[!ht]
\caption{
Simulated data from Section~\ref{subsubsec: two groups and three dimensions}. 
BIC values for the fitted models.
The best BIC value for each column is highlighted in bold, while the overall best in bold-italic.  
}
\label{tab:example 2 - BIC}
\centering
\resizebox{\textwidth}{!}{
\begin{tabular}{c c rrrrrrrrrrr}
  \toprule
  %\backslashbox{$G$}{Mixture component} 	
	$G$ &&	N	&	$t$	&	CN	&	SN	&	S$t$	&	SCN	&	SS	&	SAL	&	CSAL	\\	
	\midrule
1	&	&	-7913.407	&	-5869.537	&	-5774.226	&	-7513.308	&	-5363.720	&	-5363.330	&	-5486.237	&	-5988.705	&	-5506.787	\\
2	&	&	-6494.045	&	-3861.820	&	-3870.294	&	-6058.756	&	-3698.826	&	-3774.388	&	-3719.261	&	-4186.362	&	\textbf{\textit{-3660.907}}	\\
3	&	&	-3905.722	&	-3796.367	&	-3807.067	&	-3780.592	&	\textbf{-3671.086}	&	\textbf{-3710.395}	&	\textbf{-3701.616}	&	\textbf{-3672.434}	&	-3708.272	\\
4	&	&	\textbf{-3786.715}	&	\textbf{-3748.906}	&	\textbf{-3776.244}	&	\textbf{-3762.349}	&	-3724.686	&	-3791.960	&	-3762.420	&	-3747.273	&	-3789.211	\\
   \bottomrule
\end{tabular}
}
\end{table}
The best fitted model is the CSAL mixture with $G=2$ clusters.
For the models selected by the BIC, \tablename~\ref{tab:example 2 - ERROR and ARI} shows the corresponding ARI values with respect to the good data only. 
\begin{table}[!ht]
\caption{
Simulated data from Section~\ref{subsubsec: two groups and three dimensions}. 
ARI values, on the good data only, for the mixture models selected by the BIC.
%Bold is used for the best value of $G$ for each model, while bold-italic is used for the overall best model.  
}
\label{tab:example 2 - ERROR and ARI}
\centering
%\resizebox{\textwidth}{!}{
\begin{tabular}{c c ccccccccccc}
  \toprule
  %\backslashbox{$G$}{Mixture component} 	
	 &&	N	&	$t$	&	CN	&	SN	&	S$t$	&	SCN	&	SS	&	SAL	&	CSAL	\\	
	\midrule
$G$	&	&	4	&	4	&	4	&	4	&	3	&	3	&	3	&	3	&	2	\\
ARI	&	&	0.564 & 0.532 & 0.563 & 0.638	&	0.994 & 0.994 & 0.994 & 0.977	&	1.000	\\
   \bottomrule
\end{tabular}
%}
\end{table}
The CSAL mixture is the only model attaining a perfect classification ($\text{ARI}=1$), followed by the S$t$, SCN and SS mixtures sharing the same ARI value (0.994).
The model with the worst classification results is the $t$ mixture ($\text{ARI}=0.532$).   

\begin{figure}[!ht]
\centering
\resizebox{0.5\textwidth}{!}{
\includegraphics{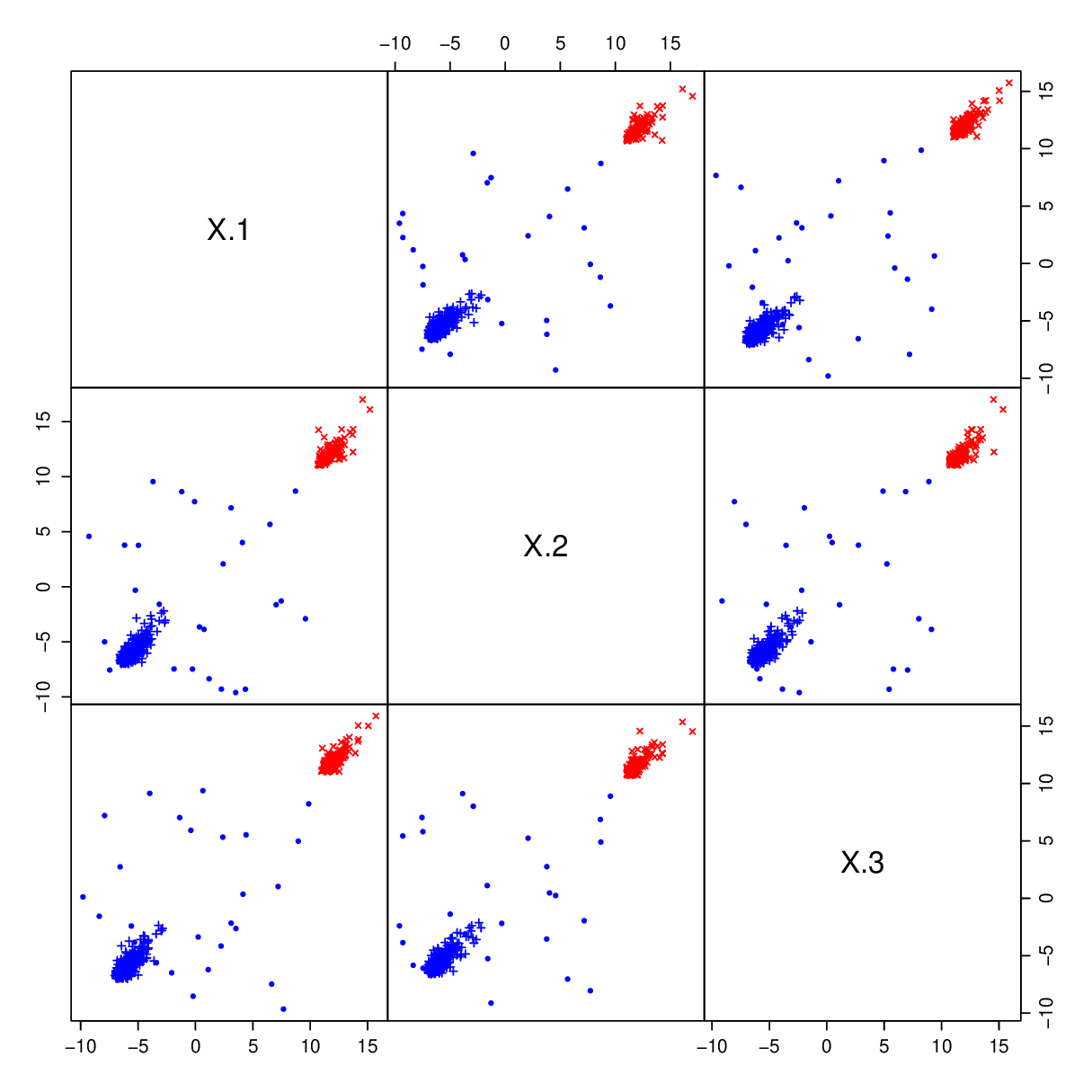} 
}
\caption{
Simulated data from Section~\ref{subsubsec: two groups and three dimensions}. 
Matrix of pairwise scatter plots with MAP-classification of the observations from the CSAL mixture with $G=2$ components.
Bullets denote detected bad points.
}
\label{fig:Outlier detection ex2}
\end{figure}

\tablename~\ref{tab:TPR and FPR ex2} reports TPRs and FPRs from the application of the outlier detection rules from the fitted CN, SCN, and CSAL mixtures with $G=2$ components.
\begin{table}[!ht]
\caption{Simulated data from Section~\ref{subsubsec: two groups and three dimensions}. 
TPRs and FPRs from the CN, SCN and CSAL mixtures with $G=2$ components.}
%\vspace{-0.1cm}
\centering{
\begin{tabular}{l ccc}
\toprule
 	  &	CN mixture & SCN mixture	&	CSAL mixture	\\	
		\midrule
TPR	&	1.000	&	1.000	&	1.000	\\
FPR	&	0.066 & 0.028 & 0.002	\\
\bottomrule
\end{tabular}
}
\label{tab:TPR and FPR ex2}
\end{table}
The competing outlier detection rules provide an optimal TPR (1.000), but the rule from the CSAL mixture is the one providing the best, almost perfect, FPR (0.002), followed by the rules from the SCN mixture ($\text{FPR}=0.028$) and CN mixture ($\text{FPR}=0.066$).
Thus, the outlier detection rule from the CSAL mixture dominates, in terms of performance, the competing rules. 
\figurename~\ref{fig:Outlier detection ex2} shows, for the CSAL mixture with $G=2$ clusters only, the matrix of scatter plots of the observations with symbols and colors diversified with respect to the group MAP-membership; detected outliers are denoted by bullets.
As we can note by comparing \figurename~\ref{fig:Outlier detection ex2} with \figurename~\ref{fig:noisescatter2}, the model is able to reproduce the underlying group-structure and to detect the background noise.
In particular, the noise is entirely captured by the bad component of the CSAL pdf placed on the bottom-left group; with reference to this component of the CSAL mixture, the estimated mixture weight is $\hat{\pi}_1=0.715$, the estimated proportion of good data is $\hat{\lambda}_1=0.925$ and the estimated degree of contamination is $\hat{\rho}_1=271.767$.
The estimated proportion of good data in the other CSAL mixture component (i.e.,~$\hat{\lambda}_2$) is approximately equal to~1.

\subsection{The bankruptcy data set}
\label{subsec:Application to real data: the bankruptcy data set}

This real data analysis considers the bankruptcy data set \citep{Altm:Fina:1968} containing, for $n=66$ manufacturing firms in the United States, the values of $p=2$ variables: the ratio of retained earnings (RE) to total assets, and the ratio of earnings before interests and taxes (EBIT) to total assets.
Half of the selected firms had filed for bankruptcy.
This data set accompanies the \textbf{ManlyMix} package \citep{ManlyMix} for \textsf{R}. 
The goal here is to predict whether a firm went bankrupt based on the two variables.
Under the same goal, this data set has been already used in the literature as an illustrative example for clustering methods assuming skewed and/or leptokurtic clusters (e.g., \citealp{Lo:Gott:Flex:2012} and \citealp{mcnicholas16a}, Chapter~7).

\begin{figure}[!ht]
\centering
\resizebox{0.4\textwidth}{!}{
\includegraphics{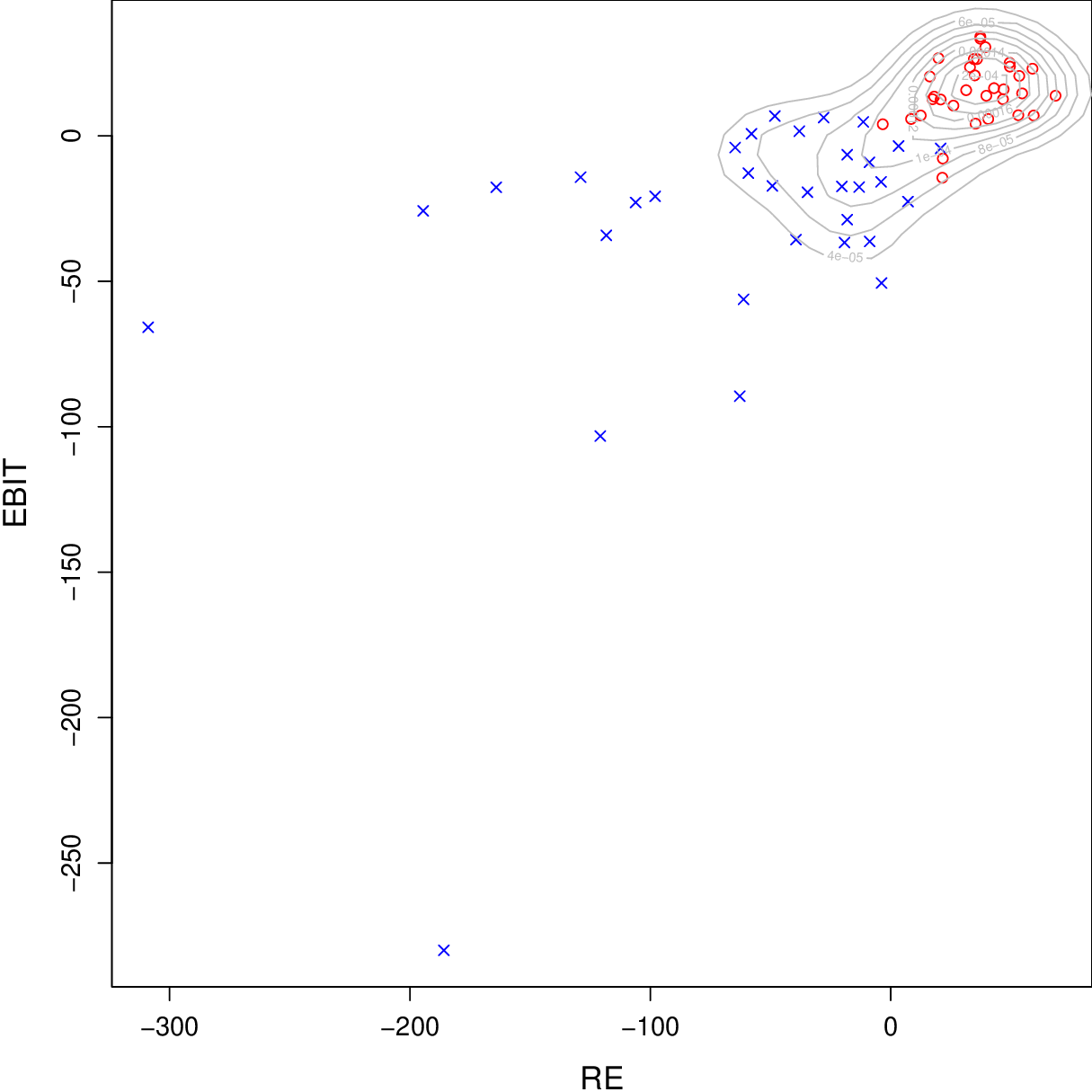} 
}
\caption{
Bankruptcy data set: scatter plot of RE and EBIT (\textcolor{black}{$\times$} denotes the solvent firms and \textcolor{red}{$\circ$} the bankrupted ones) along with contours from a bivariate normal kernel estimator.
}
\label{fig:bankruptcy scatter}
\end{figure}

\figurename~\ref{fig:bankruptcy scatter} shows the scatter plot of the data, where solvent and bankrupted firms are labeled as \textcolor{black}{$\times$} and \textcolor{red}{$\circ$}, respectively.
\begin{table}[!ht]
\caption{Bankruptcy data set: log-likelihood, BIC, and classification performance from various mixture models with $G=2$ clusters.
Bold font highlights the best value for each column.}
\label{tab:bankruptcy classification}
\centering
%\resizebox{\textwidth}{!}{
\begin{tabular}{l rrrrrr}
\toprule
Mixture components	&	\multicolumn{1}{c}{Log-likelihood}	&	\multicolumn{1}{c}{BIC}	&	\multicolumn{1}{c}{\# misclassified}	&	\multicolumn{1}{c}{Error rate}	&	\multicolumn{1}{c}{ARI}	\\	
\midrule
N	&	-652.049	&	-1350.185	&	21	&	0.318	&	0.124	\\	
$t$	&	-646.296	&	-1342.869	&	4	&	0.061	&	0.769	\\	
CN	&	-643.339	&	-1349.522	&	5	&	0.076	&	0.716	\\	
SN	&	-636.739	&	-1336.322	&	16	&	0.242	&	0.257	\\	
S$t$	&	-633.989	&	-1335.013	&	14	&	0.212	&	0.323	\\	
SCN	&	-631.301	&	\textbf{-1333.826}	&	14	&	0.212	&	0.323	\\	
SS	&	-635.874	&	-1338.782	&	15	&	0.227	&	0.289	\\	
SAL	&	-642.016	&	-1346.877	&	24	&	0.364	&	0.068	\\	
CSAL	&	-630.944	&	-1341.491	&	\textbf{3}	&	\textbf{0.045}	&	\textbf{0.824}	\\	
\bottomrule
\end{tabular}
%}
\end{table}
The contours from a bivariate normal kernel estimator, as implemented by the \texttt{bkde2D()} function of the \textbf{KernSmooth} package \citep{KernSmooth}, are also superimposed on the scatter plot.
By looking at \figurename~\ref{fig:bankruptcy scatter}, the bivariate sample is apparently bimodal, justifying the fit of two-component mixture models, with seemingly skewed clusters and possible outliers.

\tablename~\ref{tab:bankruptcy classification} presents a model comparison in terms of BIC values and shows some measures of agreement between the partition provided by each fitted model with respect to the known classification of the firms as solvent and bankrupted.
%\tablename~\ref{tab:bankruptcy classification} 

As can be seen, the SCN mixture is the best model according to the BIC, followed by the S$t$ and SN mixtures.
The CSAL mixture occupies the fifth position in this ranking.
If we take a look at the results of \tablename~\ref{tab:bankruptcy classification} from a classification point of view, the ranking changes a lot.
The worst classification result is obtained for the SAL mixture (with 24 misclassified firms), while the best one is attained by the CSAL mixture (with only 3 misclassified firms).
Interestingly, the models based somehow on the SAL distribution occupy the extremes of this ranking; this highlights how the contaminated generalization of the SAL mixture we provide can contribute to improve the classification performance of the SAL mixture.
Note also that, the classification from the CSAL mixture is even better than the one from the models compared by \citet{Lo:Gott:Flex:2012}.
The second best model, with 4 misclassified firms, is the $t$ mixture, followed by the CN mixture with 5 misclassified firms.
The other competing models have more than 13 misclassified firms each.
Thus, the best three models have leptokurtic clusters; however, allowing for skewed clusters helps the CSAL mixture to attain better classification results.
%, the further possibility of skewed 
%Thus, the two best models have leptokurtic groups and the additional possibility of skewed clusters improves the performance of the CSAL mixture over  

\begin{figure}[!ht]
\centering
\resizebox{0.4\textwidth}{!}{
\includegraphics{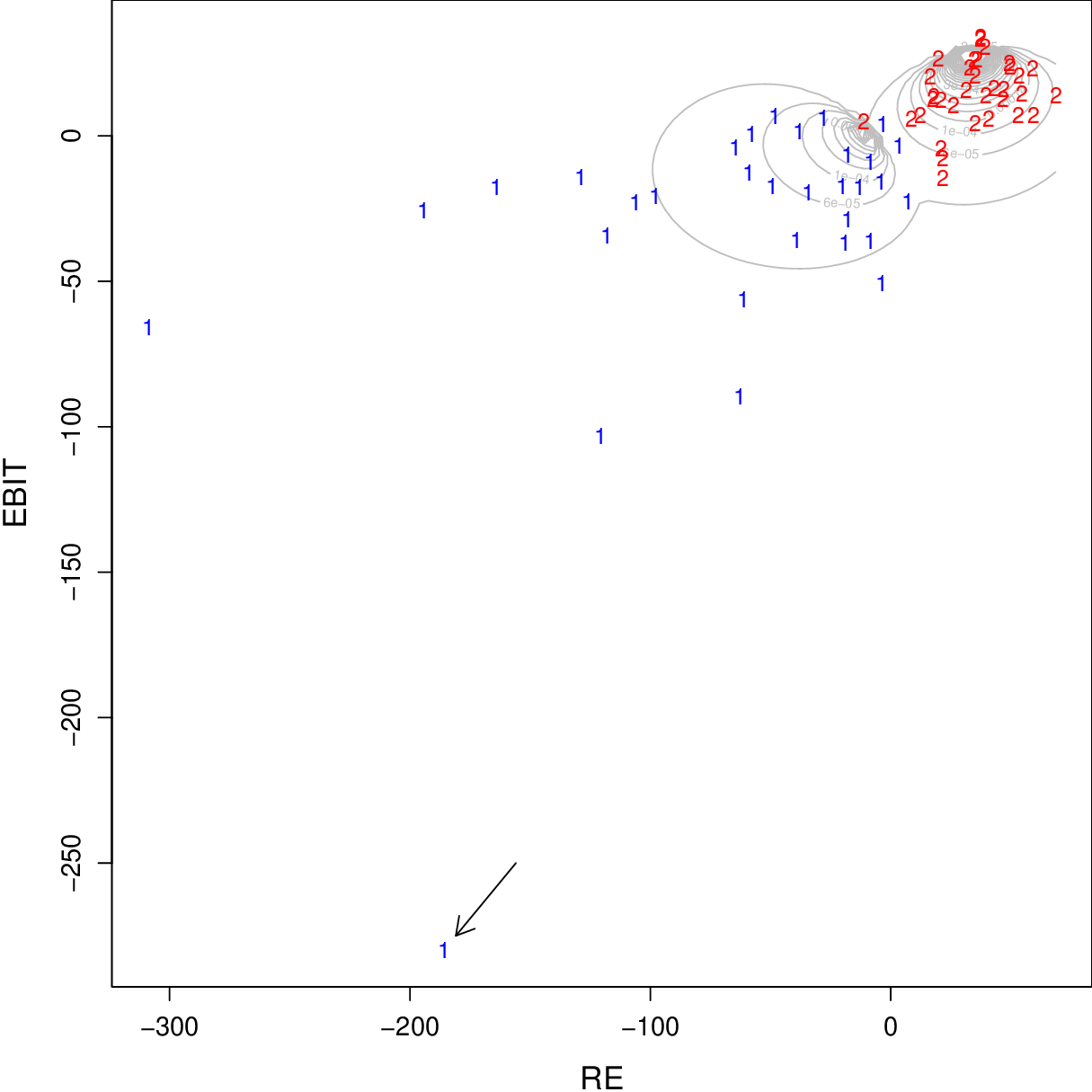} 
}
\caption{
Bankruptcy data set: scatter plot of RE and EBIT with classification and contours from the fitted two-component CSAL mixture.
The arrow indicates the unique firm detected as outlier by the model.
}
\label{fig:bankruptcy scatter CSAL}
\end{figure}

The partition from the CSAL mixture, along with its contours, is depicted in \figurename~\ref{fig:bankruptcy scatter CSAL}, where
an arrow indicates the unique outlier detected by the model; it belongs to the group of the solvent firms whose members are denoted with \textcolor{black}{$\textsf{1}$} in \figurename~\ref{fig:bankruptcy scatter CSAL}.

For the detected outlier it is also important to underline that its \textit{a~posteriori} probability to be good, in its MAP group of membership (group \textcolor{black}{$\textsf{1}$}), is practically null.
By looking at \figurename~\ref{fig:bankruptcy scatter CSAL}, there seems to be another firm, on the extreme left, which could be deemed, to eye, as an outlier, especially if one considers its distance from the bulk of the RE values.
For this point, the \textit{a~posteriori} probability to be good, in its MAP group of membership (which is group \textcolor{black}{$\textsf{1}$}), is 0.748.
This high enough probability to be good is justified if one considers the high left skewness with respect to RE induced by the first element of the estimated parameter vector $\hat{\balpha}_1=\left(-56.947,-22.096\right)^{\top}$.
Finally note that, in the group containing the outlier, the estimated proportion of good points is $\hat{\lambda}_1=0.932$ and the degree of contamination is $\hat{\rho}_1=418.872$; the so high value of $\hat{\rho}_1$ is manly due to the distance of the detected outlier from the bulk of group \textcolor{black}{$\textsf{1}$}.

%\textbf{To assess the performance of the fitted contaminated mixture, we use the values of $\hat{v}_{ig}$ to determine if a point should be assigned to the ``good'' or ``bad'' categories; specifically, we use $\text{MAP}(\hat{v}_{ig})$. This resulted in $3$ fitted groups, namely Component~1, Component~2, and Noise. 
%The observation assignments to these groups were compared against the known labels of the simulated data. Of course, it is expected that not all of these added data points will end up being recognized as bad points; in fact, some will not be bad because they will end up in the interior of a component (see \figurename~\ref{fig:salcontsim1} for example).}
%
%\textbf{As in the first scenario, not all of the added data points should end up being classified as outliers (or otherwise bad points) because not all will actually be bad. }
%
%\emph{\textbf{\figurename~\ref{fig:salcontsim2} and \figurename~\ref{fig:skewcontsim1} show the results of the clustering with contaminated SAL mixtures  for scenario two and the contaminated SN mixtures, respectively. Colour indicates the estimated group membership and shape indicates the known membership. The ${+}$ symbol denotes the simulated outliers which, as mentioned earlier, are not all actually outliers (or otherwise bad points). Note that simulated outliers that are classified as good points are assigned to a component as indicated by colour in these figures.}}

\section{Conclusions}
\label{sec:conclusion}

This paper introduced mixtures of contaminated shifted asymmetric Laplace (CSAL) distributions as a model-based clustering method for handling asymmetric clusters under the presence of outliers. 
Each component (cluster) of the mixture of CSAL distributions is a two-component SAL mixture in which one of the components, with a large prior probability, represents the ``good'' observations, and the other, with a small prior probability, with the same mode and an inflated covariance matrix --- obtained by including a contamination factor altering the skewness and scale parameters --- represents the ``bad'' observations. 
Advantageously, each mixture component is unimodal.
The ECM algorithm was employed to classify an observation by first determining its group membership, and then establishing whether it is an outlier (or otherwise bad point) within that group.
The method was applied to simulated and real data where it yielded good results when outliers (artificial or natural) were present. 
The procedure was benchmarked against other well-established mixture models, and it outperformed them, with respect to various aspects, in most of the data sets considered. 

\textcolor{black}{As an open point for further research, following the idea of \citet{McLa:Peel:Bean:Mode:2003}, \cite{mcnicholas08}, and \citet{Sube:Punz:Ingr:McNi:Clus:2013} for mixtures using the normal distribution, \citet{McLa:Bean:Jone:Exte:2007}, \citet{Andr:McNi:Exte:2011}, and \citet{Sube:Punz:Ingr:McNi:Clus:2015} for mixtures using the $t$ distribution, \citet{Punz:McNi:Robu:2014} for contaminated normal mixtures, and \citet{Murr:Brow:McNi:Mixt:2014} for mixture using the skew-$t$ distribution, parsimony, but also dimension reduction, could be introduced into the model by exploiting local factor analyzers; this would lead to mixtures of CSAL factor analyzers.
As a further possibility for further work, it would be of interest to extend the current approach to accommodate missing values in the fashion, for instance, of \citet{Lin:Lee:Ho:Onfa:2006,Lin:Ho:Shen:Comp:2009} and \citet{Lin:Lear:2014}.}

\section*{Acknowledgements}

{\small This work was supported by an Ontario Graduate Scholarship (Morris), an Early Researcher Award from the Government of Ontario (McNicholas), and a Discovery Grant from the Natural Sciences and Engineering Research Council of Canada (McNicholas). 
This work was partly supported by the Canada Research Chairs program (McNicholas). 
The computing equipment used was provided through a Research Tools and Instruments Grant from NSERC. 
The authors are grateful to Professor Tsung-I Lin for providing {\sf R} code that implements some TN calculations.

}
\end{document}